\documentclass[12pt,hyper,letterpaper]{JHEP3}
\pdfoutput=1
\usepackage[pdftex]{graphicx}
\usepackage{epsfig}
\usepackage{latexsym}
\usepackage{amssymb}
\usepackage{amsmath}
\usepackage{wrapfig}
\usepackage{boxedminipage}
\usepackage{setspace}
\usepackage{subfigure}
\usepackage{bbm}

\DeclareSymbolFont{AMSb}{U}{msb}{m}{n}
\DeclareMathSymbol{\IN}{\mathbin}{AMSb}{"4E}
\DeclareMathSymbol{\IZ}{\mathbin}{AMSb}{"5A}
\DeclareMathSymbol{\IR}{\mathbin}{AMSb}{"52}
\DeclareMathSymbol{\Q}{\mathbin}{AMSb}{"51}
\DeclareMathSymbol{\II}{\mathbin}{AMSb}{"49}
\DeclareMathSymbol{\IC}{\mathbin}{AMSb}{"43}
\DeclareMathSymbol{\IP}{\mathbin}{AMSb}{"50}
\DeclareMathSymbol{\IH}{\mathbin}{AMSb}{"48}
\DeclareMathSymbol\IA{\mathalpha}{AMSb}{"41}
\DeclareMathSymbol\IS{\mathalpha}{AMSb}{"53}
\def\N{{\cal N}}

\title{ \LARGE Mesons from global Anti-de Sitter space}
\author{Johanna Erdmenger and Veselin Filev \footnotemark[1]
\\Max-Planck-Institut f\"{u}r Physik (Werner-Heisenberg-Institut)
\\ F\"{o}hringer Ring 6, 80805 M\"{u}nchen, Germany}

\footnotetext[1]{E-mail addresses: \email{jke, filev  @mppmu.mpg.de}}

\date{\today}

\abstract{In the context of gauge/gravity duality, 
we study both probe D7-- and probe D5--branes in global Anti-de
Sitter space. The dual field theory is $\N=4$ theory on $\mathbbm{R}
\times S^3$
with added flavour. The branes undergo a geometrical phase transition in this
geometry as function of the bare quark mass $m_q $ in units of $1/R$
with $R$ the $S^3$ radius. The
meson spectra are obtained from fluctuations of the brane probes. First, we study
them numerically for finite quark mass through the phase transition.
Moreover, at zero quark mass we calculate the meson spectra
analytically 
both in supergravity and in free field theory on $\mathbbm{R}
\times S^3$ and find that the
results match: For the chiral primaries, 
the lowest level is given by the zero point energy or by
the scaling dimension of the operator corresponding to the
fluctuations, respectively. The higher levels are equidistant.  
Similar results apply to the descendents. 
Our results confirm the physical interpretation that  the
mesons cannot pair-produce any further 
when their zero-point energy exceeds their binding
energy. 
}

\preprint{MPP-2010-159\\DIAS-STP-10-13}

\begin{document}
\maketitle

\section{Introduction}

The AdS/CFT correspondence in its original form involves the
near-horizon limit of D3--branes. This limit leads to the $AdS_5 \times
S^5$ geometry, where the Anti-de Sitter factor of the geometry is
realized as the Poincar\'e patch of $AdS_5$. 

Following \cite{Karch:2002sh}, there have been extensive studies of probe D7--branes
 in this geometry which on the dual field theory side
leads to added hypermultiplets in the fundamental representation of
the gauge group. For preserving $\N=2$ supersymmetry, the probe D7--brane has
to wrap a subspace which is asymptotically $AdS_5 \times S^3$ near the
boundary. The complete meson spectra arising from the fluctuations of the probe
brane, ie. for all scalar, fermion and
vector modes, have been found in \cite{Kruczenski:2003be}. For instance, for
scalar mesons the result is \cite{Kruczenski:2003be}
\begin{equation} \label{Myers}
M_s (n,l) =  \frac{2L}{R^2} \sqrt{(n+l+1) (n+l+2)} \,
,  \qquad
\frac{L}{R^2} = \sqrt{2} \pi \frac{m_q}{\sqrt \lambda} \, ,
\end{equation}
where $L$ is the embedding coordinate of the D7 brane, $R$ is the AdS
radius, $m_q$ is the quark mass and $\lambda$ the 't Hooft
coupling. $n$ is the main quantum number and $l$ is the quantum number
of the $SO(4)$ symmetry associated with the $S^3$ asymptotically
wrapped by the D7--brane probe near the boundary. The spectrum has a
large degeneracy since it depends only on the combination
$(n+l)$. This is expected from the $\N=2$ supersymmetry of the
system. For general Dp/Dq--brane systems, similar spectra were found in ref.~\cite{{Arean:2006pk},{Myers:2006qr}}.

By embedding a D7--brane probe into deformed versions of  $AdS_5 \times
S^5$, physical phenomena may be described. Examples are 
chiral
symmetry breaking, which is obtained, by embedding a D7--probe into a confining background 
\cite{Babington:2003vm}, and
a first-order
meson-melting phase transition at finite temperature which arises when
embedding a
probe brane into the AdS-Schwarzschild black hole geometry
\cite{Babington:2003vm,Kirsch:2004km, Kruczenski:2003uq, Hoyos:2006gb}.  
For the AdS-Schwarzschild geometry, a topology-changing phase
transition occurs, depending on the ratio of quark mass over
temperature, between branes that reach the black hole horizon and
those that do not.

Supersymmetric embeddings of D5--brane probes wrapping an 
$AdS_4 \times S^2$ of $AdS_5 \times S^5$ have first been studied in 
\cite{DeWolfe:2001pq}. In this case, the dual field
theory is a superconformal `defect' theory \cite{Erdmenger:2002ex} 
in which the additional hypermultiplets
are confined to a (2+1)-dimensional subspace.

In \cite{Karch:2009ph} (see also \cite{Karch:2006bv}), the authors
investigate the embedding of  D7-- and D5--brane probes into global AdS. Global AdS is
dual to $\N=4$ Super Yang-Mills theory on $\mathbbm{R} \times S^3$.
Moreover these authors study the thermodynamical properties of these
systems by considering brane probes in global thermal AdS, where also 
the time direction is compactified, such that the field theory is
defined on $S^1 \times S^3$. Again in these geometries they find a
topology-changing phase transition, depending on the ratio of quark
mass over $S^3$ radius,  between those embeddings that reach
the $S^3$ and those that do not. Using critical exponents, the authors show (see also refs.~\cite{Frolov:2006tc, Mateos:2006nu, Filev:2008xt}) that the phase transition is third order for D7--brane probes,
while it is first order for D5--brane probes. They also consider
fluctuations of the brane and calculate the mass of the lowest-lying
scalar meson mode. The authors find that the spectrum has a kink at the phase
transition. Moreover, by making use of the Polyakov loop, they study
the physical properties of the low-energy phase. In this phase, the
zero-point energy of the mesons -- which is due to the finite volume of
the $S^3$ -- is larger than their binding energy. This means in
particular that the mesons are deconfined in the sense that they
cannot pair-produce any more. String breaking is no longer possible. 

In this paper we investigate the mesonic spectrum   on $\mathbbm{R}
\times S^3$ in further
detail. We look at scalar and vector fluctuations for both the D7-- and
the D5--brane probe cases. For the D7 brane probe, we establish the
complete bosonic spectrum on the gravity side. 
Our most important result concerns the
spectrum at vanishing quark mass, for which we perform analytical
calculations both on the gravity and on the field theory side. On the
gravity side, we map the fluctuation equations of motion to equations
of Schr\"odinger type. For the chiral primaries, 
we find that the energy of the ground state is
given by the dimension of the operator dual to the fluctuations. The
higher fluctuation modes, labelled by the quantum number $n$, 
are equidistant. For the descendents, we also find an equidistant
spectrum. However the energy of the ground state is no longer equal to
the dimension of the operator. 

We then turn to the field theory side, and in particular to 
the free $\N=2$ theory in $3+1$ dimensions obtained by adding a flavour
hypermultiplet to the original $\N=4$ theory. We expand the fields on
the $S^3$ within $\mathbbm{R} \times S^3$.
The spectrum of the mesonic
composite operators is obtained by combining the expansions of the
component fields, making use of the appropriate Clebsch-Gordan
coefficients. 
This requires 
a careful analysis of
the transformation properties of the mesonic composite
operators under the antipodal map.\footnote{The antipodal map $A : S^n
  \rightarrow S^n$, 
defined by $A(x) = - x$, sends every point on the sphere to its
antipodal point.} 
For both the chiral primaries and the descendents, we
find the same result for the spectrum as in the gravity calculation.
As an example, let us quote our result for scalar D7--brane mesons, 
which is
\begin{equation} \label{examplemass}
M(n,l,\tilde l)= \frac{1}{R} (3+ 2n + l + \tilde l) \, .
\end{equation}
Here, $R$ is the radius of the $S^3$ in the field theory directions,
which gives rise to the conformal mass $1/R$. Note that since this
mass arises from the background metric rather than from the D7--brane
boundary conditions, (\ref{examplemass}) is independent of the
't Hooft coupling $\lambda$, while \eqref{Myers} is ${\cal O} 
(1/\sqrt{\lambda})$. Moreover,
\eqref{examplemass} depends on the main quantum number $n$, as well as
on the $S^3$ quantum numbers $l$ and $\tilde l$, where $l$ refers to
the internal $S^3$, asymptotically wrapped by the D7--brane, while
$\tilde l$ refers to the $S^3$ in $\mathbbm{R} \times S^3$ in the field
theory directions. The lowest mode, ie. the zero-point energy, is given by the conformal
dimension of the dual operator, $\Delta=3$. 

It is remarkable that the free field calculation and the gravity
calculation of the meson spectrum agree. 
This is of course due to non-renormalization theorems which hold 
also in $\N=2$ theory. From the physics perspective 
this confirms the physical interpretation, already advocated in 
\cite{Karch:2009ph}, that the mesons cannot pair-produce even at
strong coupling when confined into a small volume such that their
zero-point energy is larger than their binding energy. 

We note that when setting $\tilde l$ to zero, our new result
\eqref{examplemass}, which in this case depends on the combination
$2n+l$, is less degenerate than \eqref{Myers} which depends on $n+l$. 
This linked to the fact that for supersymmetric field theories on $S^3$,
scalars, fermions and vectors in the same multiplet have different
conformal mass. Also, the supersymmetry transformations of the field
theory fermions are modified to contain curvature-dependent terms. 
For $\N=4$ Super-Yang-Mills theory on $\mathbbm{R} \times S^3$, these issues have been 
studied in detail in \cite{Kim:2003rza}. We expect similar results to
hold for the $\N=2$ theory considered here.

Our calculation shows - in a simple example - 
that it is possible to directly compare gravity and
field theory calculations in top-down approaches to holographic models
of physical relevance. Such top-down models have been used widely
recently to holographically describe physical phenomena such as
superfluidity, quantum critical points and transport properties
\cite{superconductors1,superconductors2,critical,conductivity}. 
These models have the advantage - as compared to
bottom-up models - that the dual field theory is explicitly known.
We hope that further field-theory studies will follow soon for a more
detailed comparison between the weak and strong coupling aspects of a
given model.

This paper is organized as follows. In section 2 we introduce the
general setup of global AdS and its probe brane embeddings, and
present the phase transitions for the D7 and D5 brane cases. In
section 3 we study the probe brane fluctuations as function of the
bare quark mass and discuss the behaviour of the meson spectrum for
both the phase transition and the case of vanishing quark mass. In
section 4, we analytically compute the meson spectra at zero quark
mass on the gravity side. We obtain the full bosonic spectrum for the
D7 brane fluctuations, as well as some characteristic examples for the
D5 brane case. In section 5 we present the field theory calculation
at zero quark mass and show that it agrees with the gravity results.
We end with concluding remarks in section 6.

\section{General Setup}
Our starting point is  AdS$_5\times S^5$ in global coordinates:
\begin{equation}
ds^2=-(1+w^2/R^2)d t^2+w^2d\Omega_3^2+\frac{dw^2}{1+w^2/R^2}+R^2d\Omega_5^2\ . \label{Glob}
\end{equation}
It is convenient to introduce the following radial coordinate:
\begin{equation}
u=\frac{1}{2}(w+\sqrt{R^2+w^2})\ .
\end{equation}
The metric (\ref{Glob}) in these coordinates is given by:
\begin{equation}
ds^2=-\frac{u^2}{R^2}\left(1+\frac{R^2}{4u^2}\right)^2dt^2+\frac{u^2}{R^2}\left(1-\frac{R^2}{4u^2}\right)^2d\Omega_3^2+\frac{R^2}{u^2}\left(du^2+u^2d\Omega_5^2\right)\ .\label{Globu}
\end{equation}
Note that $u\ge R/2$. In this way the transverse $\IR^6$ has a ball of radius R/2 sitting at the origin. 
\subsection{Probe D7--brane.}
\begin{figure}[htbp] 
   \centering
   \includegraphics[width=4.5in]{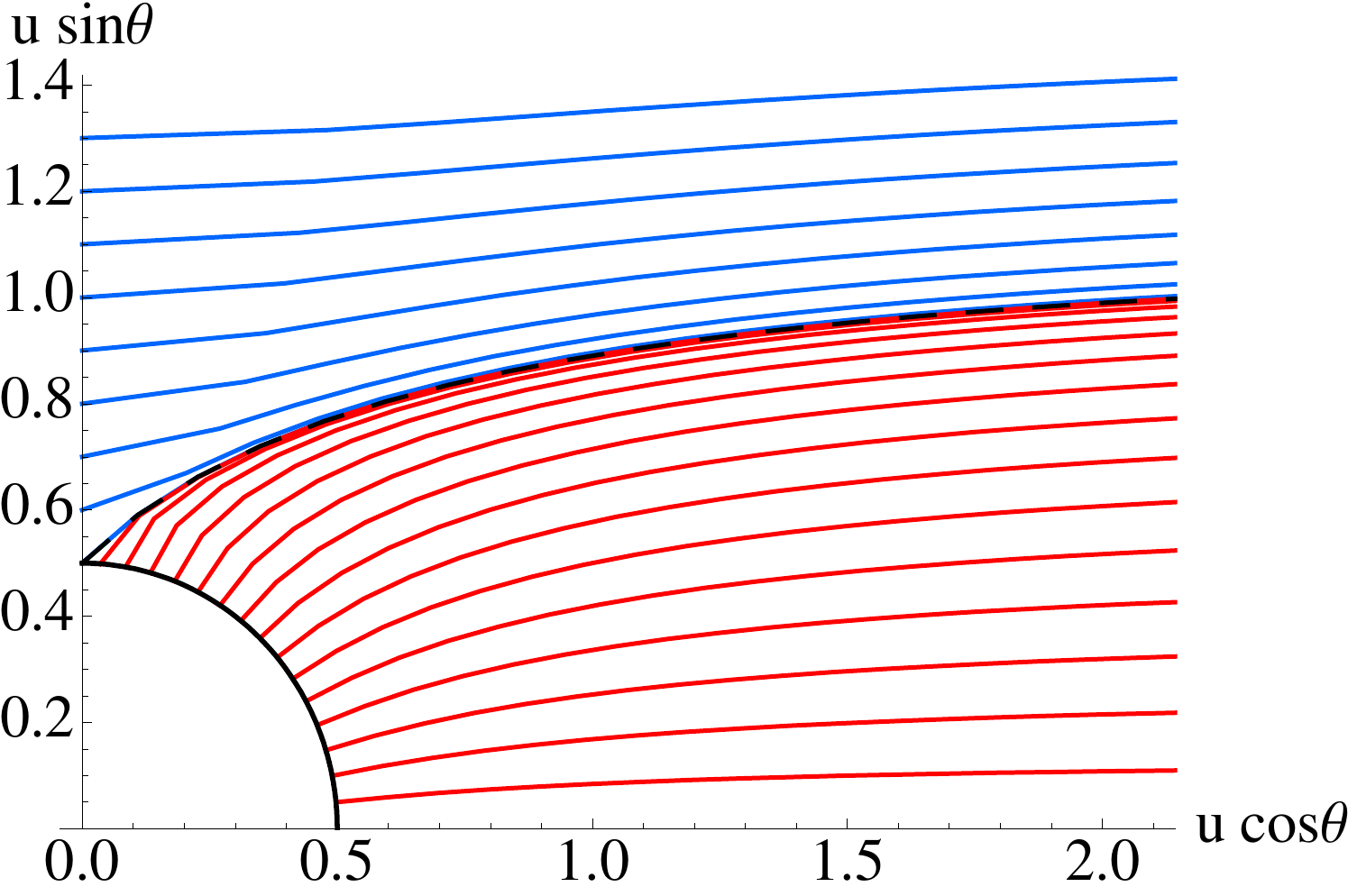} 
   \caption{\small Blue curves correspond to Minkowski embeddings and Red curves correspond to ``ball" embeddings. The black dashed line corresponds to the critical embedding.}
   \label{fig:fig1}
\end{figure}

Let us introduce a probe D7--brane. To this end it is convenient to write the metric of unit $S^5$ in the following coordinates:
\begin{equation}
ds_{S^5}^2=d\theta^2+\cos^2\theta d\Omega_3^2+\sin^2\theta d\phi^2\ .
\end{equation}
Now if we let the D7-brane be extended along the $AdS_5$ part of the geometry and wrap an $\tilde S^3\subset S^5$, the radial part of the corresponding DBI lagrangian is given by:
\begin{equation}
{\cal L} \propto \left(1-\frac{R^4}{16u^4}\right)\left(1-\frac{R^2}{4u^2}\right)^2u^3\cos^3\theta\sqrt{1+u^2\theta'(u)^2}\ .\label{lagr}
\end{equation}
Possible embeddings split into two classes Minkowski embeddings and ``ball" embeddings correspondingly wrapping shrinking $S^3$ cycles in the $S^5$ and AdS$_5$ parts of the background \cite{Karch:2006bv} (look at figure \ref{fig:fig1}). The two classes are separated by a critical embedding which has conical singularity at the ball (represented by the dashed curve in figure \ref{fig:fig1}). 
According to the AdS/CFT dictionary the bulk dynamics of the scalar $\theta(u)$ encodes the dynamics of the dual gauge invariant operator. In particular one can read off the source and the {\it vev} of the operator from the asymptotic behaviour of $\theta(u)$. In our case the operator is the fundamental bilinear and the source and {\it vev} of the operator correspond to the mass of the hypermultiplet and the fundamental condensate. The precise dictionary has been derived in ref.~\cite{Karch:2005ms}, where an elegant renormalization prescription has been offered. Let us briefly review the results of refs.~\cite{Karch:2005ms,Karch:2006bv} in a slightly modified form relevant to our notations.
For large $u$ the solution to the equation of motion derived from (\ref{lagr}) has the following expansion:
\begin{equation}
\theta(u)=\frac{\theta_0R}{u}+\frac{\theta_2R^3}{u^3}-\frac{\theta_0R^3}{2u^3}\log{\frac{u}{R}}+\dots\ .
\end{equation}
After following the renormalization prescription outlined in ref.~\cite{Karch:2005ms} the condensate of the theory can be calculated in terms of the parameters $\theta_0,\theta_2$. For our choice of radial coordinate the result is given by:
\begin{equation}
\langle\bar\psi\psi\rangle\propto -2\theta_2+\frac{\theta_0^3}{3}+\theta_0\log\theta_0;\ .
\end{equation}
If we split the $\IR^6$ space corresponding to $du^2+u^2d\Omega_5^2$ to $\IR^4\times \IR^2$ and define radial coordinates $\rho=u\cos\theta$ and $L=u\sin\theta$ the DBI lagrangian in this coordinates is given by:
\begin{equation}
{\cal L}\propto \left(1+\frac{R^2}{4u^2}\right)\left(1-\frac{R^2}{4u^2}\right)^3\rho^3\sqrt{1+L'^2}\ .\label{DBIL}
\end{equation}

 The profile of the D7--brane embedding $L(\rho)$ has the following asymptotic behaviour at large $\rho$:
\begin{equation}
L(\rho)=m+\frac{c_1}{\rho^2}-\frac{m}{2\rho^2}\log\rho+\dots\ .
\end{equation}
Further more one has the relations $m=\theta_0R$ and $c_1=R^3(\theta_2-\theta_0^3/6)$. The condensate of the theory is then given by:
\begin{equation}
\langle\bar\psi\psi\rangle\propto-2c_1+mR^2\log(m/R)\equiv -2c\ ,
\end{equation}
where we have defined a new parameter $c$ proportional to the condensate. Note also that the bare mass of the hypermultiplet is proportional to $m$ the exact relation is $m_q=m/2\pi\alpha'$. 

\begin{figure}[htbp] 
   \centering
   \includegraphics[width=4.5in]{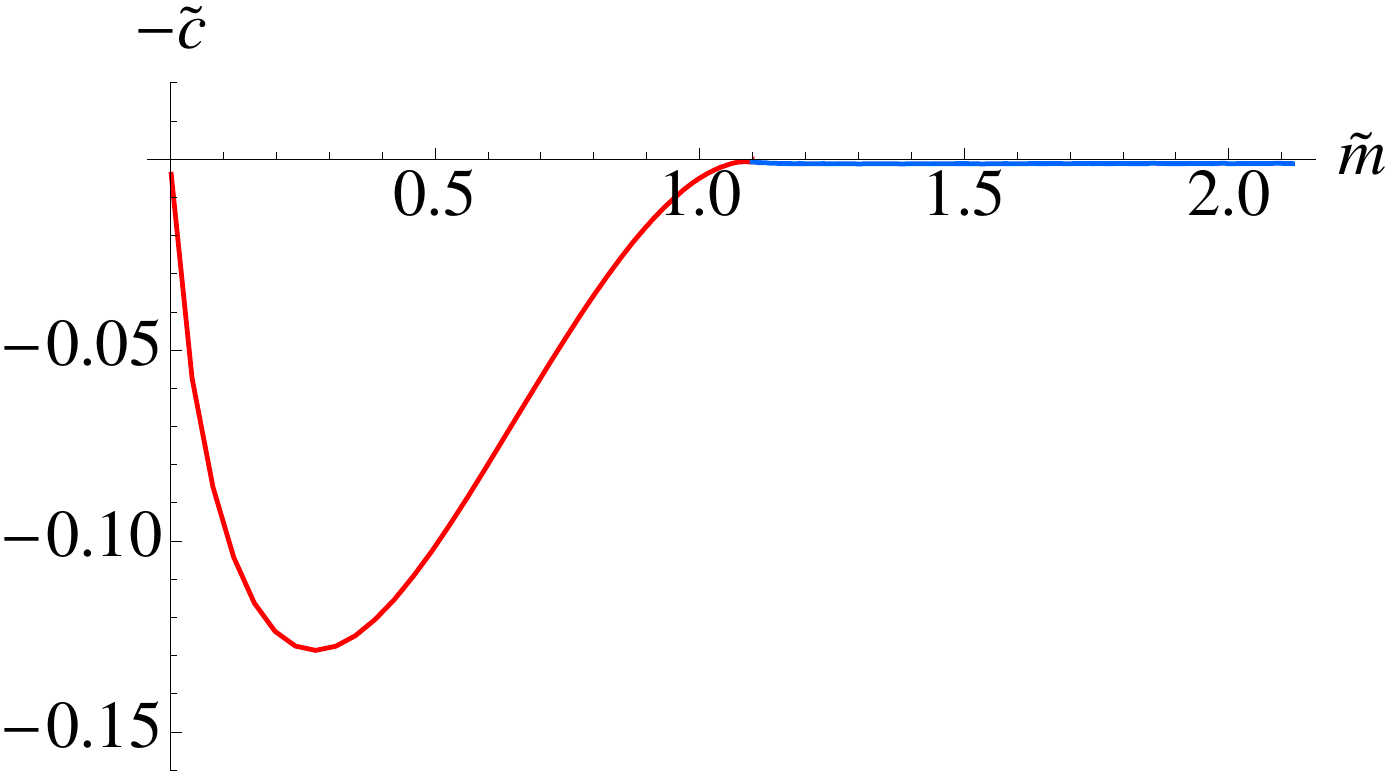} 
   \caption{\small A plot of the condensate $-\tilde c$ versus the bare mass $\tilde m$. The states corresponding to ``ball" embeddings are presented by the red curve and the states corresponding to Minkowski embeddings are presented by the blue curve.}
   \label{fig:fig2}
\end{figure}

After solving numerically for the D7--brane embeddings we can generate a plot of the equation of state $c(m)$ presented in figure \ref{fig:fig2}, where we have used dimensionless parameters $\tilde m=m/R$ and $\tilde c=c/R^3$.
As one can see there is no apparent multi-valued region near the
transition from Minkowski to ``ball" embeddings. In fact it has been
shown that the phase transition is continuous and is of a third order \cite{Karch:2009ph}. 

Let us comment briefly on the physical meaning of the parameter $\tilde m=m/R$ and in particular its $R$ dependence. Consider a constant $u=\Lambda\gg R$ slice of the AdS$_5\times S^5$ space-time. The induced metric has the asymptotic form:
\begin{equation} 
ds_{\Lambda}^2=\frac{\Lambda^2}{R^2}(dt^2+R^2d\Omega_3)+R^2d\Omega_5^2 \ .\label{induced_metric_Lambda}
\end{equation}
Here $\Lambda$ is specifying the energy scale of the dual gauge theory. It is natural to identify the radius of the AdS$_5$ space-time $R$ with the radius of the three sphere $R_3$ where the holographically dual field theory is defined. Furthermore we know that $m=2\pi\alpha' m_q$, where $m_q$ is the bare mass of the hypermultiplet. However the quantity $m_q/R_3$ is obviously not a dimensionless parameter. To clarify this let us consider a rescaling $\Lambda\to\gamma\Lambda$ of the energy scale. The metric in equation (\ref{induced_metric_Lambda}) becomes:
\begin{equation}
ds_{\gamma\Lambda}=\frac{\Lambda^2}{R^2}(\gamma^2dt^2+\gamma^2R^2d\Omega_3^2)+R^2d\Omega_5^2
\, .
\end{equation}
We can always scale the time coordinate $t\to t/\gamma$ to compensate for the rescaling of the energy scale. Note that this suggests that the bare mass of the hypermultiplet should also be rescaled $m_q\to\gamma m_q$. Furthermore the radius of the three sphere is now $R_3=\gamma R$ and hence we can write: $\gamma=R_3/R$.  For the parameter $\tilde m$ we obtain:
\begin{equation}
\tilde m=\frac{\gamma m_q(2\pi\alpha')}{R}=m_q R_3\frac{2\pi\alpha'}{R^2}=\frac{\pi}{\sqrt{2}}\frac{m_qR_3}{\sqrt{\lambda}}\ ,\label{physmeaning}
\end{equation}
where we have used that $R^2$ is related to the t'Hooft coupling of the dual gauge theory via: $R^2=2\lambda\alpha'$. Equation (\ref{physmeaning}) specifies the physical meaning of the parameter $\tilde m$. Now we can interpret the phase transition as taking place at constant bare mass $m_q$ and varying radius of $S^3$. Note that varying $R_3$ corresponds to varying the Casimir energy of the dual gauge theory and hence it is  essentially a quantum phase transition. 
\subsection{Probe D5--brane}
Let us review the case of a probe D5--brane studied in ref.~\cite{Karch:2009ph}. To this end it is convenient to write the $S^5$ part of the geometry in the following coordinates:
\begin{equation}
ds_{S^5}^2=d\psi^2+\cos^2\psi d\Omega_2^2+\sin^2\psi d\tilde\Omega_2^2\ .
\end{equation}
Next we define: $r=u\cos\psi$ and $l=u\sin\psi$. The metric (\ref{Globu}) can then be written as:
\begin{eqnarray}
ds^2&=&-\frac{u^2}{R^2}\left(1+\frac{R^2}{4u^2}\right)^2dt^2+\frac{u^2}{R^2}\left(1-\frac{R^2}{4u^2}\right)^2[d\alpha^2+\sin^2\alpha(d\beta^2+\sin^2\beta
d\gamma^2)]\nonumber \\ \label{parmD5}
&+&\frac{R^2}{u^2}[dr^2+r^2 d\Omega_2^2+dl^2+l^2d\tilde\Omega_2^2]\ .
\end{eqnarray}
Now if we let the D5--brane be extended along the $t,\beta,\gamma,r,\Omega_2$ directions and has a non-trivial profile along $l$, the corresponding DBI lagrangian is given by:
\begin{equation}
{\cal L}\propto \left(1+\frac{R^2}{4u^2}\right)\left(1-\frac{R^2}{4u^2}\right)^2r^2\sqrt{1+l'^2}\ .\label{DBID5}
\end{equation}
It is easy to show that the solution to the equation of motion has the following asymptotic behaviour at large $r$:
\begin{equation}
l(r)=m+\frac{c}{r}+\dots.
\end{equation}
Note that unlike the D7--brane case there is no extra logarithmic term  \cite{Karch:2005ms} in the asymptotic expansion of $l(r)$. One can then directly relate the coefficient $c$ to the fundamental condensate of the theory ($\langle\bar q q\rangle\propto -c$). It is convenient to define dimensionless quantities $\tilde c=c/R^2$ and  $\tilde m=m/R$. 

\begin{figure}[htbp] 
   \centering
   \includegraphics[width=3.2in]{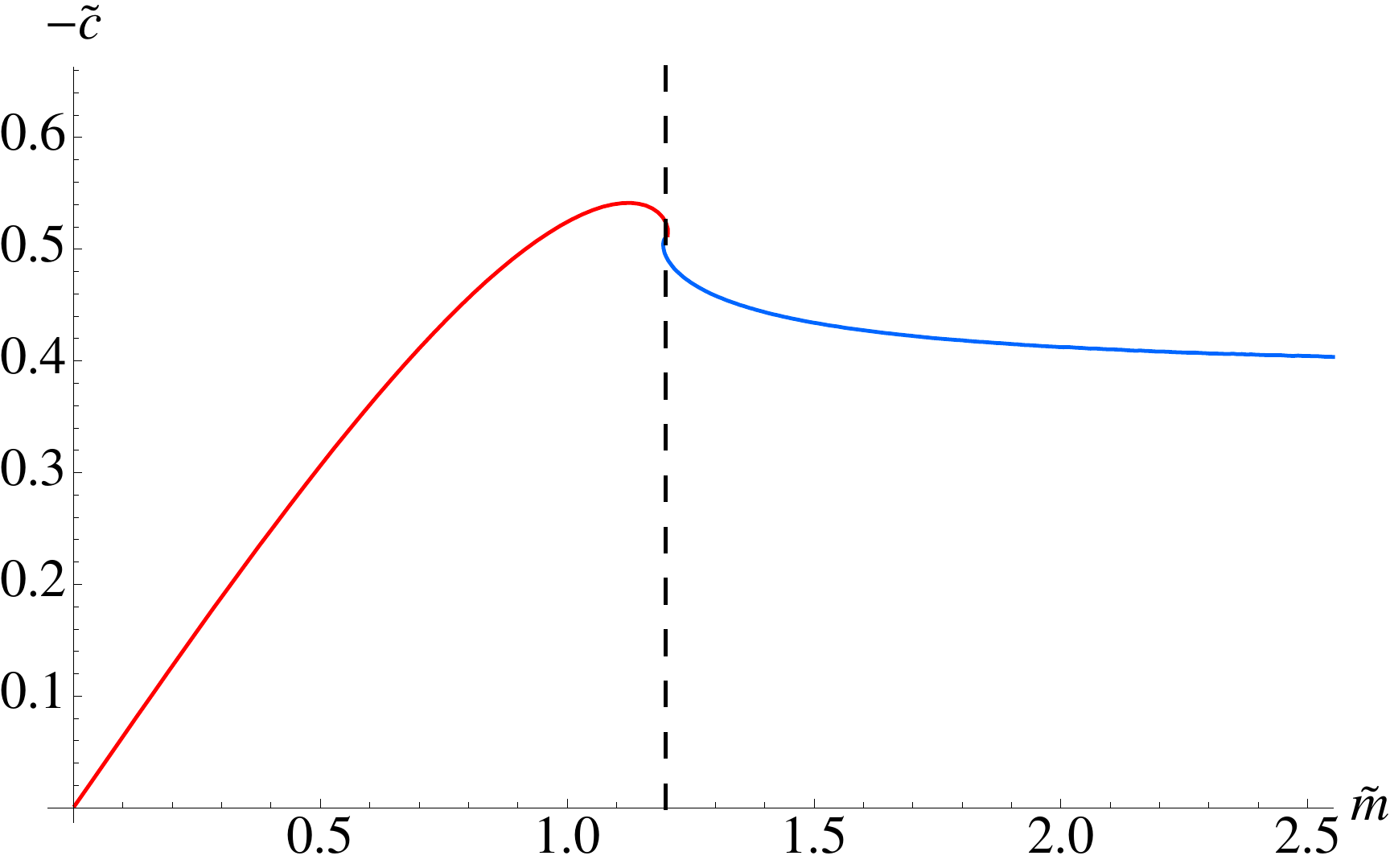} 
     \includegraphics[width=3.2in]{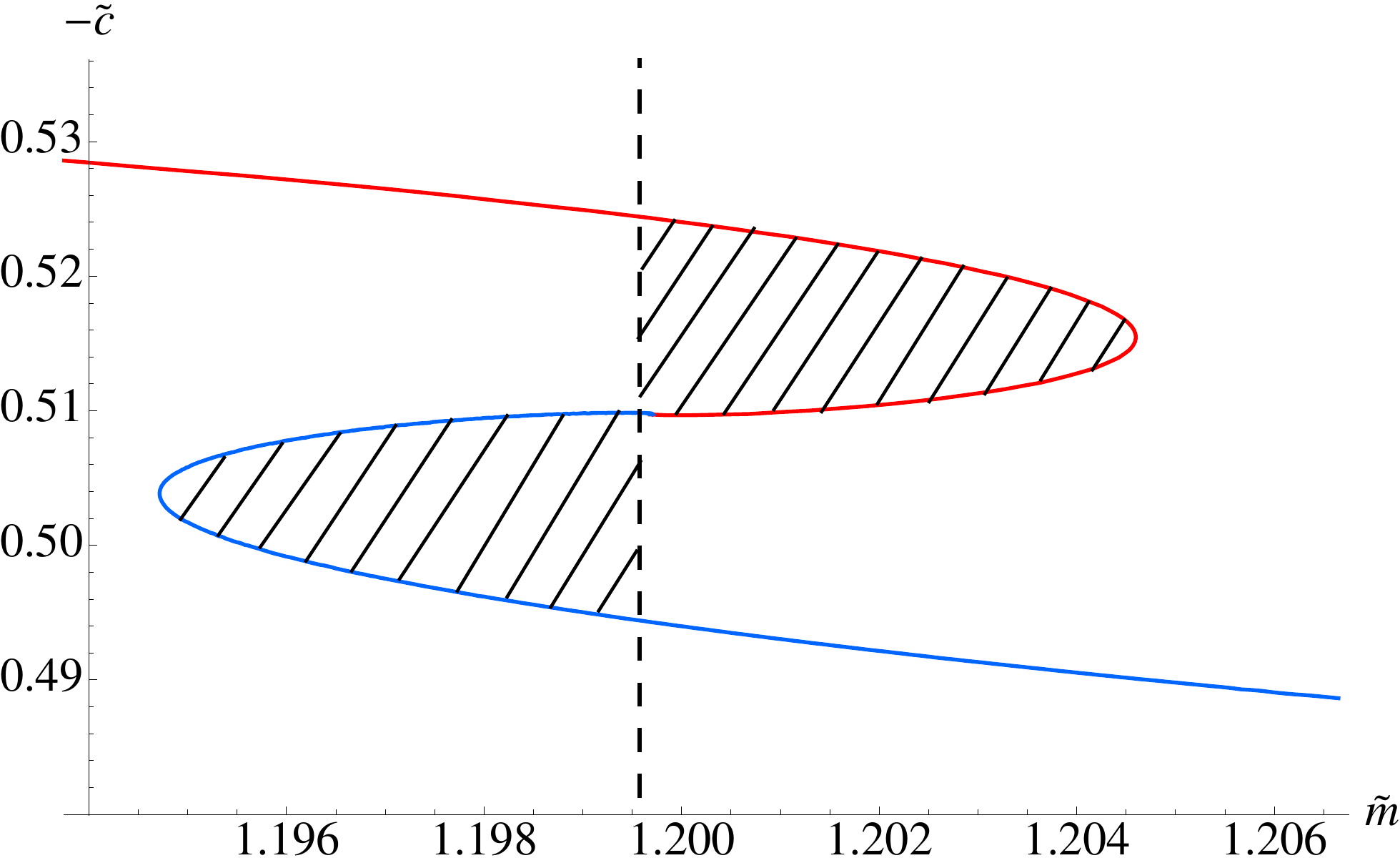} 
   \caption{\small A plot of the condensate $-\tilde c$ versus the
     bare mass $\tilde m$ for the D5--brane probe. The states
     corresponding to 
``ball" embeddings are presented by red curves and the states corresponding to Minkowski embeddings are presented by blue curves. The zoomed in plot suggests a first order phase transition in the dual gauge theory.}
   \label{fig:condD5}
\end{figure}

After solving numerically the equation of motion of the probe D5--brane we obtain the plot of $-\tilde c$ versus $\tilde m$ presented in figure \ref{fig:condD5}.
As one can see there is a multi--valued region near the state corresponding to the critical embedding. This suggests that in this case the corresponding phase transition in the dual gauge theory is a first order one. The value of the critical bare mass at which the phase transition takes place can be obtained using the {\it equal area law} as illustrated in the second plot in figure  \ref{fig:condD5}. The authors of ref.~\cite{Karch:2009ph} showed that the order of the phase transition can be inferred analytically, by calculating appropriate critical exponents. 

\section{Meson spectra}
\label{sec:mesonspectra}

Some aspects of the  spectrum corresponding to fluctuations of the D7--brane
embeddings were studied in ref.~\cite{Karch:2009ph}, where the lowest
mode for each of the fluctuations was considered. In particular, the authors 
considered the scalar meson frequency squared corresponding to the fluctuation
along the profile of the embedding (the coordinate L in our case) and found
that as function of the quark mass parameter  $\tilde m$, it has a kink
precisely at the critical value $\tilde m_*$ where the phase transition
occurs. This kink was shown to be consistent with the fact that the phase
transition is third order.
On the other hand, the spectrum of fluctuations along the coordinate $\phi$
was shown to be a smooth function of $\tilde m$ across the phase transition. 
Moreover, the authors of ref.~\cite{Karch:2009ph} pointed out 
that for large $\tilde m \gg 1$ (ie. for large  $S^3$ radius) 
the spectrum matches with the meson spectrum of the D3/D7 brane intersection 
obtained in ref.~\cite{Kruczenski:2003be}.

The authors of ref.~\cite{Karch:2009ph} explain the behaviour of the lowest
meson mode as follows: Consider the case that at fixed quark mass, the system
is squeezed into smaller and smaller volume. When the compactification radius
$R$ passes through its critical value $R^*$, the mesons are expected to
deconfine as their finite volume zero point energy within becomes larger
than their binding energy. Beyond the phase transition the theory is in a
deconfined phase, in the sense that it cannot pair--produce.

Below  we provide a more detailed analysis of the spectrum of meson-like
excitations by considering higher excited modes ($n>0$ in the notations of
ref.~\cite{Kruczenski:2003be}). Our study shows that at the phase transition,
the qualitative behaviour of the higher modes 
is the same as that of the ground state, as far as the appearance of the
kink is concerned. For values $\tilde m < \tilde m^*$, we continue to
find a discrete spectrum. In particular,
for the limit of zero bare mass ($\tilde m=0$), we are able to perform an
analytic calculation, presented in detail in section 4 and 5, 
mapping the fluctuation equation of motion to an
equation of Schr\"odinger type, and find that the 
discrete spectrum is equidistant. 
Moreover, for the fluctuations dual to chiral primaries
we find that the energy of the ground state 
is given by the conformal dimension of the dual operator (in units of $1/R$). 
A similar structure emerges for
the fluctuations of a D5 brane probe.

Moreover, for the D5 brane probe fluctuations 
we also investigate the $\tilde m$ dependence 
and show that the qualitative behaviour near the critical embedding is
consistent with the phase transition being first order. In particular, the
spectrum develops tachyonic states exactly at those points $\tilde m_\infty$ of the multi-valued
region of the condensate $\tilde c (\tilde m)$ where the slope of the condensate diverges.  

We also turn on the Kaluza--Klein modes of the D7--brane in the $S^5$ part
of the geometry corresponding to operators with non-vanishing R--charge. We
find that the large degeneracy of the spectrum of the D3/D7 system considered
in ref.~\cite{Kruczenski:2003be} is lifted for the theory on $S^3$. In the limit of large $\tilde m$ (large radius of $S^3$) this degeneracy is restored. In the limit of zero bare mass the degeneracy is only partially restored. Furthermore, once again the energy of the ground state is determined by the engineering dimension of the operator ($\Delta =3+l$) and the spectrum is equidistant.

The outline of our calculations and results in this section is as follows:
We first present our numerical results for fluctuations of a probe D7---brane
along the coordinate $L$. We describe the qualitative behaviour across the
phase transition and discuss the structure of the spectrum at vanishing bare
mass. Next we present our numerical results for the spectrum of fluctuations
along $\phi$ for the D7--brane case, discussing also the Kaluza--Klein
modes. In the limit of vanishing bare quark mass, we compare 
our numerical results to the analytical results obtained below in
section 4 and 5. Moreover we analyze the spectrum of fluctuations of the D5--brane. We discuss both fluctuations along the $l$ coordinate and along the Neumann--Dirichlet coordinate (the polar coordinate of $S^3\subset AdS_5$).

\subsection{Fluctuations of the D7--brane embedding.} 

\subsubsection{Fluctuations of the transverse scalars.}
In order to study the light meson spectrum, we look for
the quadratic fluctuations of the D7--brane embedding along the
transverse directions parametrized by $L,\phi$. To this end we
expand:
\begin{equation}
L=\bar L+2\pi\alpha'\delta L;~~~\phi=2\pi\alpha'\delta\phi;\ ,\end{equation}
in the lagrangian (\ref{DBIL}) and leave only terms of order $(2\pi\alpha')^2$. After some calculations we obtain the following lagrangian for the quadratic fluctuations along $L$:
\begin{equation}
 {\cal
 L}_{LL}^{(2)}\propto\frac{1}{2}\sqrt{-g}G_{LL}\frac{g^{\alpha\beta}}{1+L'^2}\partial_{\alpha}\delta
 L\partial_{\beta}\delta
 L+\frac{1}{2}\left[\partial_{L}^2\sqrt{-g}-\frac{d}{d\rho}\left(\frac{L'}{1+L'^2}\partial_L\sqrt{-g}\right)\right]\delta
 L^2\, , \label{LagrFlD7}
\end{equation}
where $G_{LL}$ is the corresponding component of the metric of the
$AdS_5\times S^5$ background and $g^{\alpha\beta}$ is the inverse of the
induced metric on the worldvolume of the D7--brane. Next we obtain the
equation of motion for $\delta L$ and consider an ansatz $\delta L=e^{i\omega
  t}h(\rho)$. Finally we solve the resulting differential
equation for $h(\rho)$ numerically and require normalizability of the solution, which
leads to a quantized spectrum. The corresponding spectrum is presented in figure \ref{fig: SpectrumLD7}, 
\FIGURE{ 
\centering
\includegraphics[width=4.5in]{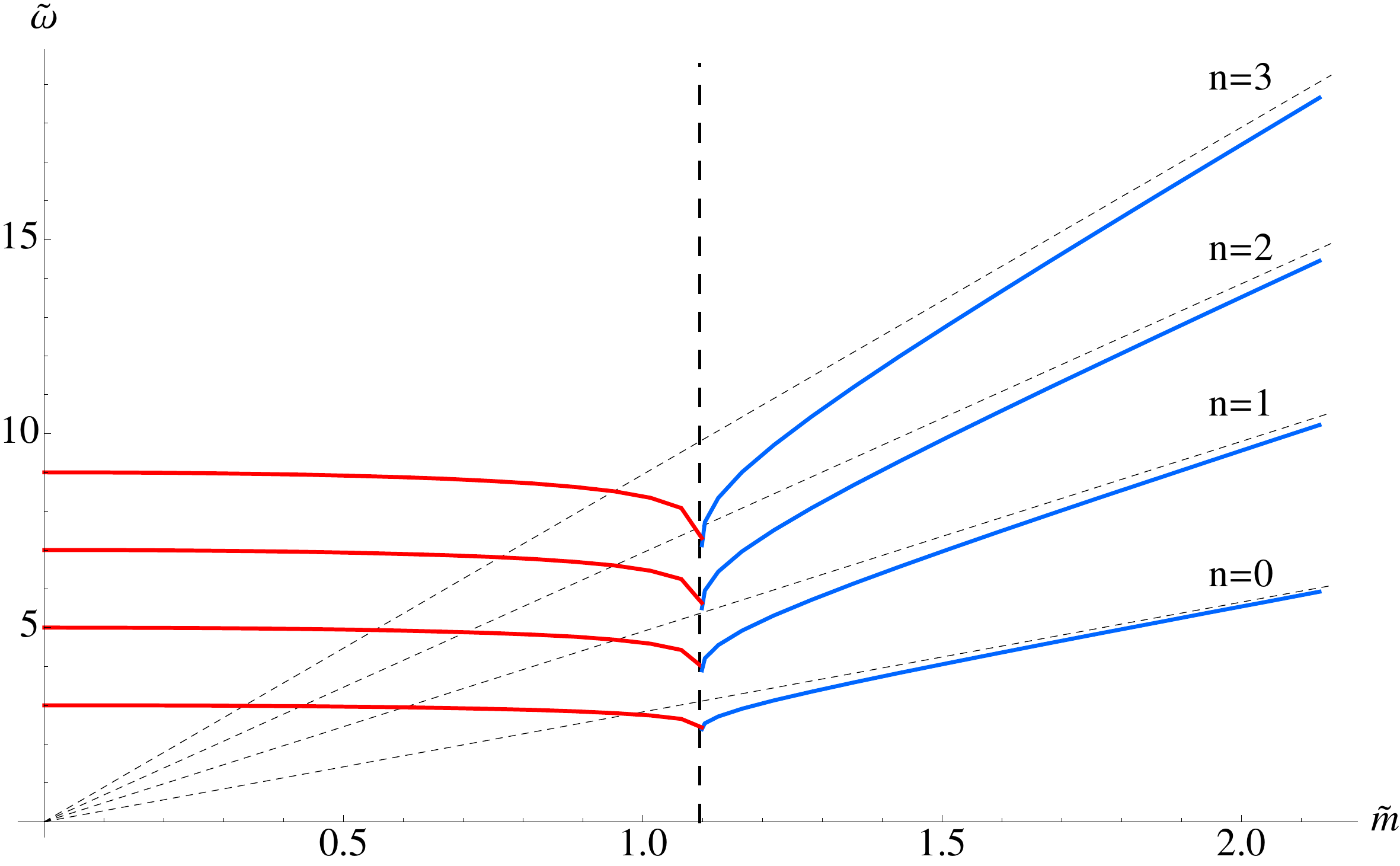} 
\caption{\small A plot of the meson mass $\tilde\omega$ versus the bare quark mass $\tilde m$ for the spectrum of D7 fluctuations along $L$. The black dashed lines correspond to the spectrum of D3/D7 intersection \cite{Kruczenski:2003be} given by equation (\ref{flatD3_D7}), one can observe the good agreement at large $\tilde m$. At the critical bare mass (represented by the vertical dashed line) the spectrum has a kink. Interestingly at $\tilde m=0$ the spectrum is discrete and equidistant taking only integer values given by equation (\ref{Spectm=0D7}).}
\label{fig: SpectrumLD7}
}
where the dimensionless quantity $\tilde \omega=\omega R$ has been defined. At large values of $\tilde m$ (large radius of $S^3$) the spectrum approaches the spectrum of the D3/D7 intersection given by
\begin{equation}
\omega=\frac{2m}{R^2}\sqrt{(n+1)(n+2)}\, . \label{flatD3_D7}
\end{equation}
At the critical bare mass (represented by the vertical dashed line in figure
\ref{fig: SpectrumLD7}) the spectrum has a kink as pointed out in
ref.~\cite{Karch:2009ph}.  Below the phase transition the spectrum has a discrete quasi--equidistant structure which becomes exact at vanishing bare mass:
\begin{equation}  
\omega=(2n+3)\frac{1}{R}\ . \label{Spectm=0D7}
\end{equation}
Note that in \eqref{Spectm=0D7}, the energy of the ground state ($n=0$)
is given by the engineering dimension $\Delta=3$ of the operator dual to the
fluctuations considered. 

As the authors of ref.~\cite{Karch:2009ph} argue in this phase the
mesons are deconfined. The equidistant structure of the spectrum then
arises from  the Kaluza-Klein tower of the collection of fields on $\tilde S^3$ corresponding to the field content of the deconfined mesons.  
We refer the reader to sections \ref{sec:gravity}  and \ref{sec:fieldtheory} for analytic derivations of
equation (\ref{Spectm=0D7}) both on the supergravity and on the 
field theory side of the correspondence.

 Note that the dependence of the spectrum on the 't Hooft coupling $\lambda$ is through the parameter $\tilde m$. According to equation (\ref{flatD3_D7}), at large bare mass $\tilde m$ the dimensionless quantity $\tilde\omega$ depends linearly on the parameter $\tilde m$  and hence in view of equation (\ref{physmeaning}) is proportional to $1/\sqrt{\lambda}$. In agreement with figure~\ref{fig: SpectrumLD7}, below the phase transition, i.e. at small values of $\tilde m$,  the spectrum changes weakly with $\tilde m$. Asymptotically, for $\tilde m\to 0$ the spectrum becomes independent of $\tilde m$ and hence of $\lambda$. 

Let us now focus on the spectrum of fluctuations along $\phi$. After expanding
to second order in $\alpha'$ in the DBI lagrangian (\ref{DBIL}) and
substituting the ansatz $\delta \phi=e^{i\omega t}Y_{lm}(S^3)f(\rho)$ into the resulting equation of motion we obtain:
\begin{equation}
\frac{1}{\sqrt{-g}\bar L^2}\partial_{\rho}\left(\sqrt{-g}\bar L^2\frac{f'(\rho)}{1+{\bar L}'^2}\right)+\left(\frac{R^4\omega^2}{(u^2+\frac{R^4}{4})^2}-\frac{l(l+2)}{\rho^2}\right)f(\rho)=0\ .
\label{EOMPhi}\end{equation}
Note that the quantum number $l$ represents Kaluza--Klein modes on $S^3\subset S^5$ and labels representations of the global SU(2) R--symmetry of
the dual gauge theory. Let us first focus on the spherically symmetric case
($l=0$). Solving equation (\ref{EOMPhi}) numerically and requiring
normalizability of the solution, we obtain the meson mass $\tilde \omega$
versus the bare quark mass $\tilde m$ presented in figure \ref{fig:SpectPhil0}.
\FIGURE{
\centering
\includegraphics[width=4.5in]{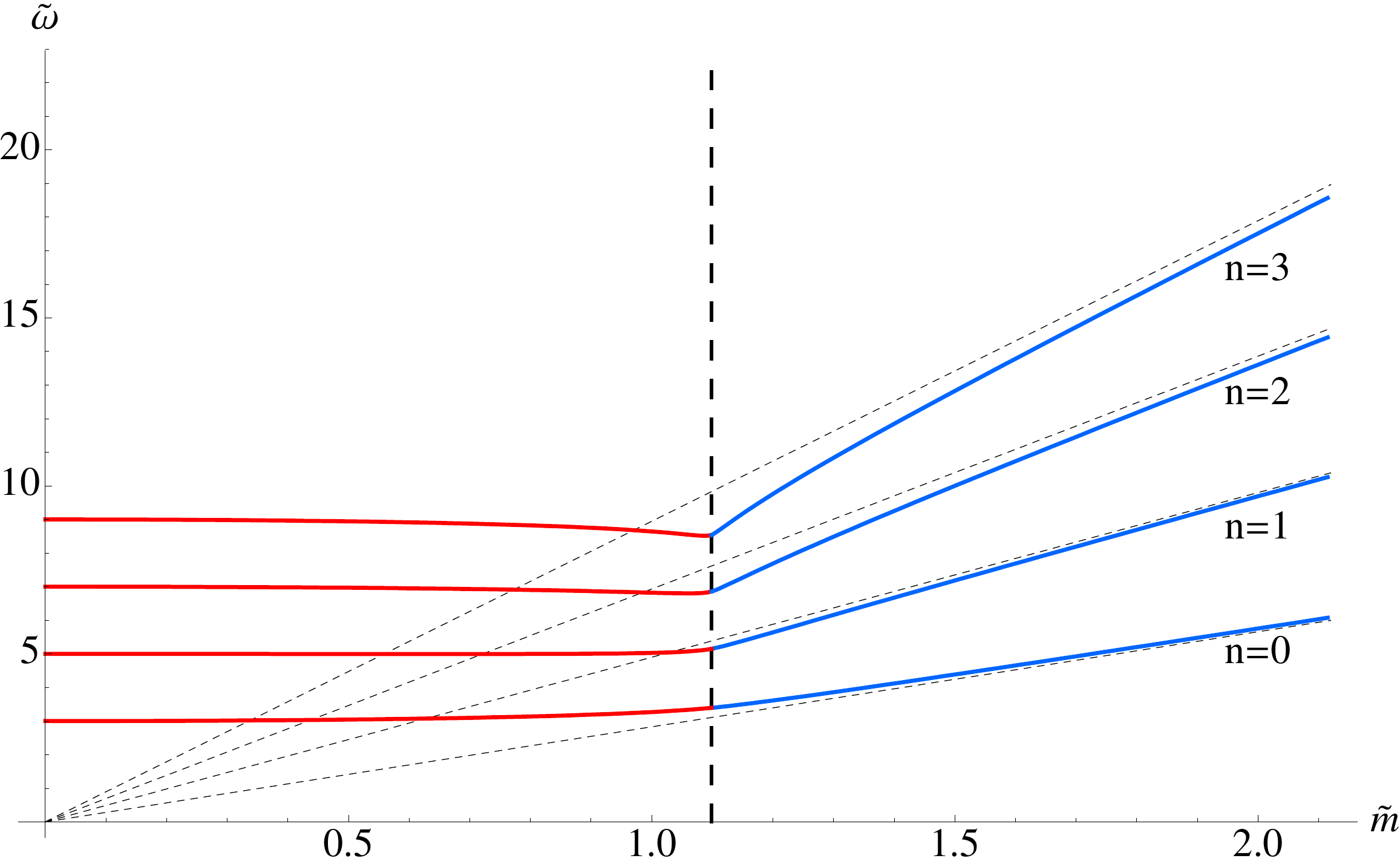} 
\caption{\small A plot of the meson mass $\tilde\omega$ versus the bare quark
  mass $\tilde m$ for the spectrum of D7 fluctuations along $\phi$. At large $\tilde m$ the spectrum is described by equation (\ref{flatD3_D7}). At zero bare mass ($\tilde m=0$) the spectrum is equidistant and is described by equation (\ref{Spectm=0D7}). Note that the spectrum is a smooth function of $\tilde m$ across the phase transition.}
\label{fig:SpectPhil0}}
We see that for large bare mass (or equivalently for large radius of $S^3$),
the spectrum matches the spectrum of the D3/D7 system described by equation
(\ref{flatD3_D7}). In the limit of zero bare mass ($\tilde m=0$),  the spectrum is equidistant and is again identical to spectrum of fluctuations along $L$ described by equation (\ref{Spectm=0D7}). At the phase transition (the vertical dashed line in figure \ref{fig:SpectPhil0}) the spectrum is a smooth function of $\tilde m$. 

Our next goal is to study the spectrum at non-zero $l$. Note that in the
``flat case" considered in ref.~\cite{Kruczenski:2003be}, ie.~with a Minkowski
boundary in Poincar\'e coordinates, the spectrum has degeneracy in $n$ and $l$ and is described by generalization of equation~(\ref{flatD3_D7}):
\begin{equation}
\omega=\frac{2m}{R^2}\sqrt{(n+l+1)(n+l+2)}\ .\label{genD3_D7}
\end{equation}
On the other hand,  the theory on $\tilde S^3$, 
which we consider here, is not supersymmetric for generic values of $\tilde m$
and we do not expect the spectrum to be degenerate. Indeed our numerical investigations confirm this. As an illustration, in figure~\ref{fig:SpectPhil} we provide the spectrum of the ground state ($n=0$) for various values of $l$.
\FIGURE{
\centering
\includegraphics[width=4.5in]{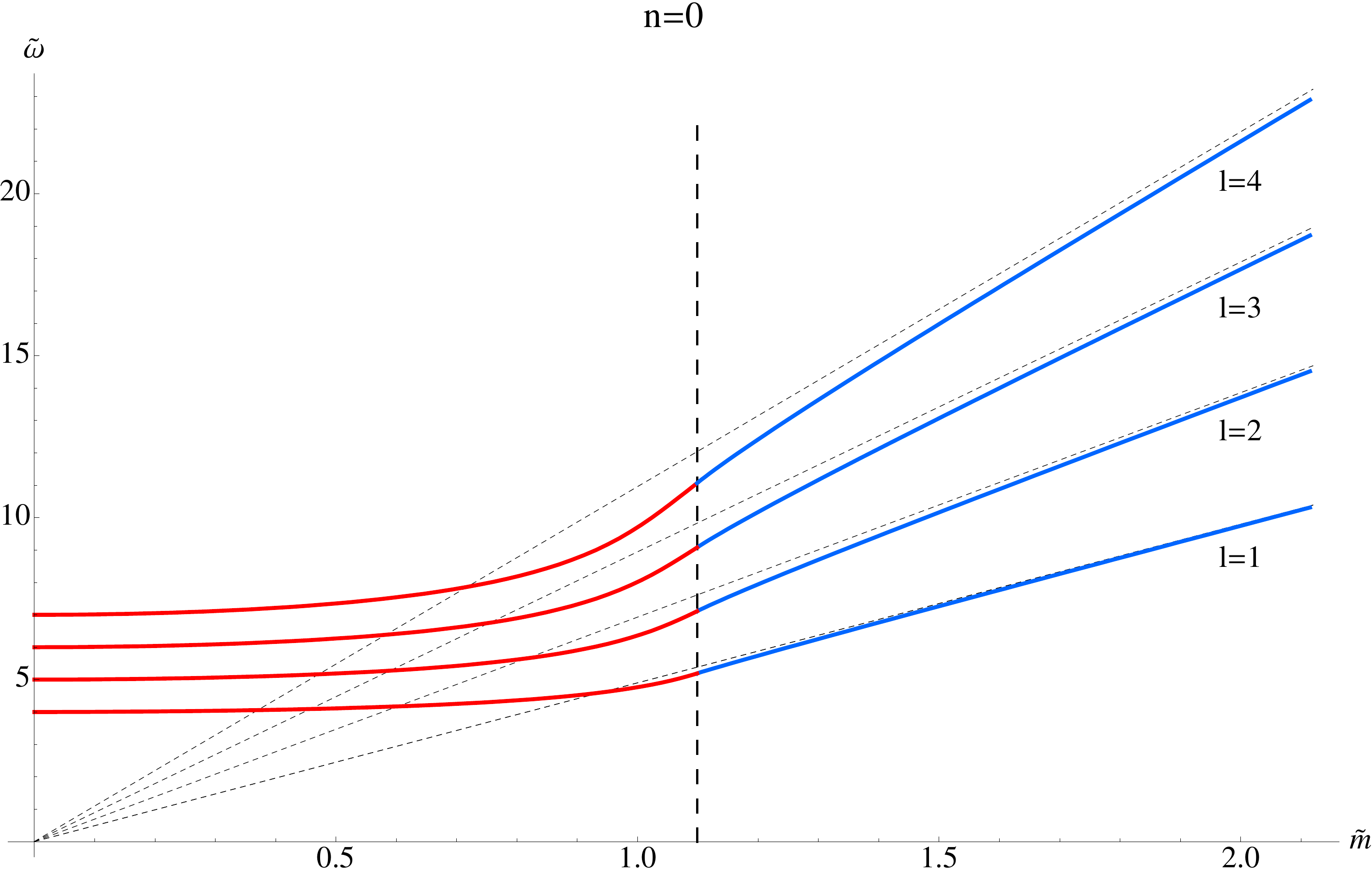} 
\caption{\small D7 brane fluctuations: A plot of the meson mass $\tilde \omega$ versus the bare quark
  mass $\tilde m$ for various values of $l$. At large $\tilde m$ the spectrum is given by equation (\ref{genD3_D7}). At zero bare mass the spectrun is described by $\tilde \omega=3+l$. }
\label{fig:SpectPhil}
}
As one may expect, at large $\tilde m$ the spectrum again matches the spectrum
of the D3/D7 system given by equation (\ref{genD3_D7}). We also compared
to the $l=0,n\neq0$ result in figure~\ref{fig:SpectPhil0} and verified that
for generic values of $\tilde m$ the spectrum is not degenerate. This is
particularly evident in the deconfined phase (the red colored
curves). Interestingly, the spectrum at $\tilde m=0$ is described by $\tilde
\omega=3+l$, which suggests that the spectrum again has some
degeneracy. This fits with the fact that supersymmetry is restored at $\tilde
m=0$. In fact, by taking  the limit $\bar L\to0$ in equation (\ref{EOMPhi})
and analyzing the resulting equation of motion analytically, we find  that the spectrum at $\tilde m$ is given by:
\begin{equation}
\omega=(3+2n+l)\frac{1}{R}\label{SpctPhinl}\ .
\end{equation}
We refer the reader to section 4 for a detailed derivation of the spectrum
at $\tilde m=0$. Notice however that the combination $3+l$ in equation
(\ref{SpctPhinl}) is equal to the engineering dimension of the corresponding
dual field theory operator. Indeed if we look at the asymptotic form of the equation of motion (\ref{EOMPhi}) for large $\rho$ we obtain:
\begin{equation}
f''(\rho)+\frac{3}{\rho }f'(\rho)-\frac{l(l+2)}{\rho^2}f(\rho)=0\ .\label{asymptEOM}
\end{equation}
The solution of equation (\ref{asymptEOM}) is of the general form:
\begin{equation}
f(\rho)=\rho^{l}\delta m+\rho^{-(l+2)}\delta c
\end{equation}
and hence according to the standard AdS/CFT dictionary, the corresponding gauge invariant operator indeed has conformal dimension $\Delta=3+l$. This suggests that for generic meson--like excitations the spectrum at zero bare mass is given by:
\begin{equation}
\omega=(\Delta+2n)\frac{1}{R}\ ,\label{SpectSug}
\end{equation}
where $\Delta$ is the conformal dimension of the field operator dual 
to the lowest fluctuation mode in agreement with the standard field/operator
dictionary. Note also the increase in units of $2n$ between the equidistant
levels.  We study the spectrum in more detail in sections 4 and 5
below. For completeness and to check the validity of equation
(\ref{SpectSug}), we now turn to analyzing the spectrum of fluctuations of a probe D5--brane
in the next subsection.

\subsection{Fluctuations of the D5--brane embedding.}

In order to study the meson spectrum corresponding to fluctuations of the
D5--brane along the coordinate $l(r)$, defined above \eqref{parmD5}, 
we expand  $l(r)=\bar l(r)+(2\pi\alpha')\delta
l(r)$ in the DBI Lagrangian (\ref{DBID5}) and leave only terms of order
$\alpha'^2$. Next we obtain the corresponding equation of motion and consider
an ansatz $\delta l(r)=e^{i\omega t}\eta(r)$. After solving the equation of motion
numerically and imposing regularity of the solution, we obtain the meson
spectrum as a function of the bare quark mass. Our results for the ground state and
the first few excited states are presented in figure \ref{fig:SpectrD5}, 
where $\tilde\omega=\omega R$ as before.
\FIGURE{
\centering
\includegraphics[width=3.3in]{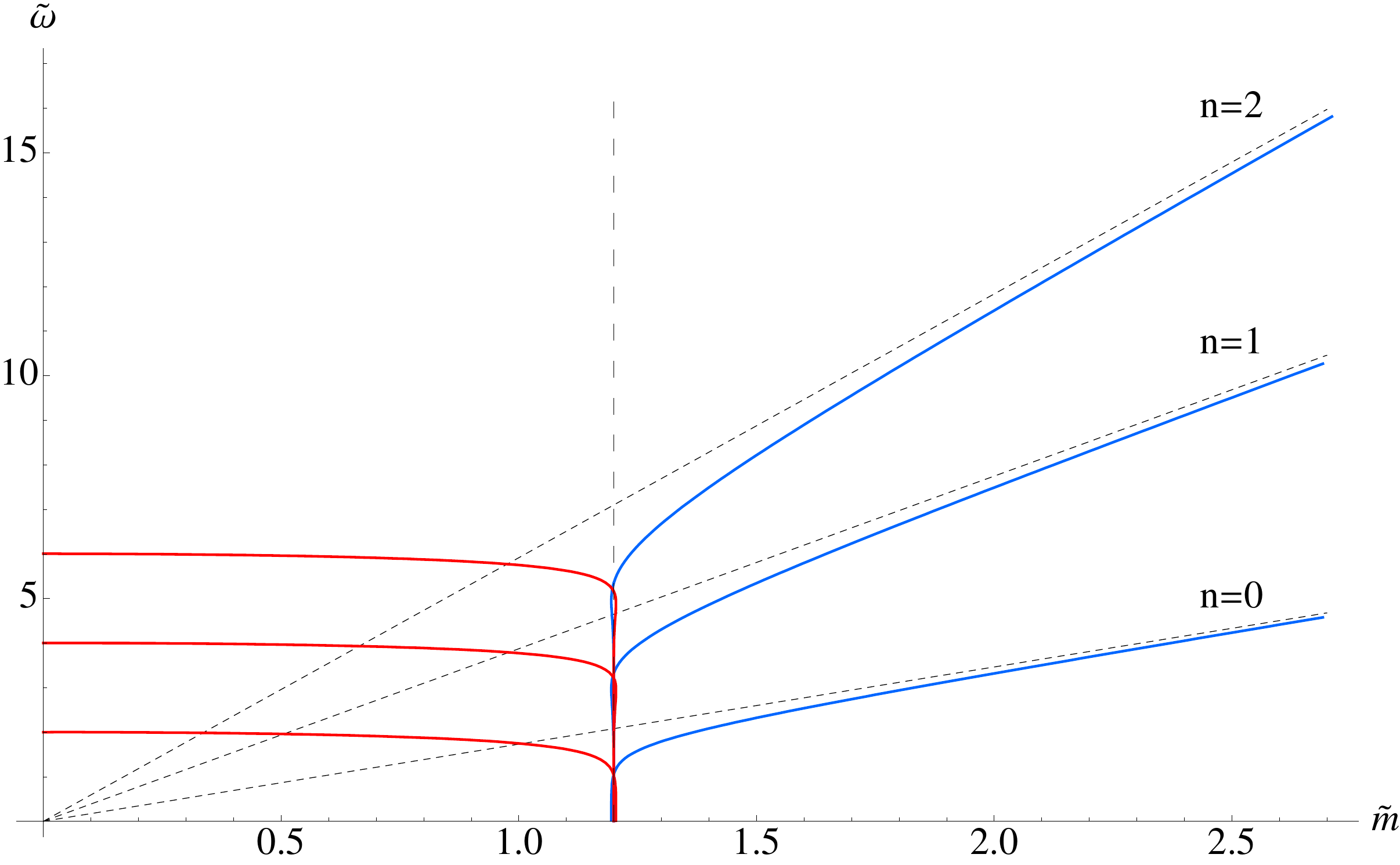} 
\includegraphics[width=3.3in]{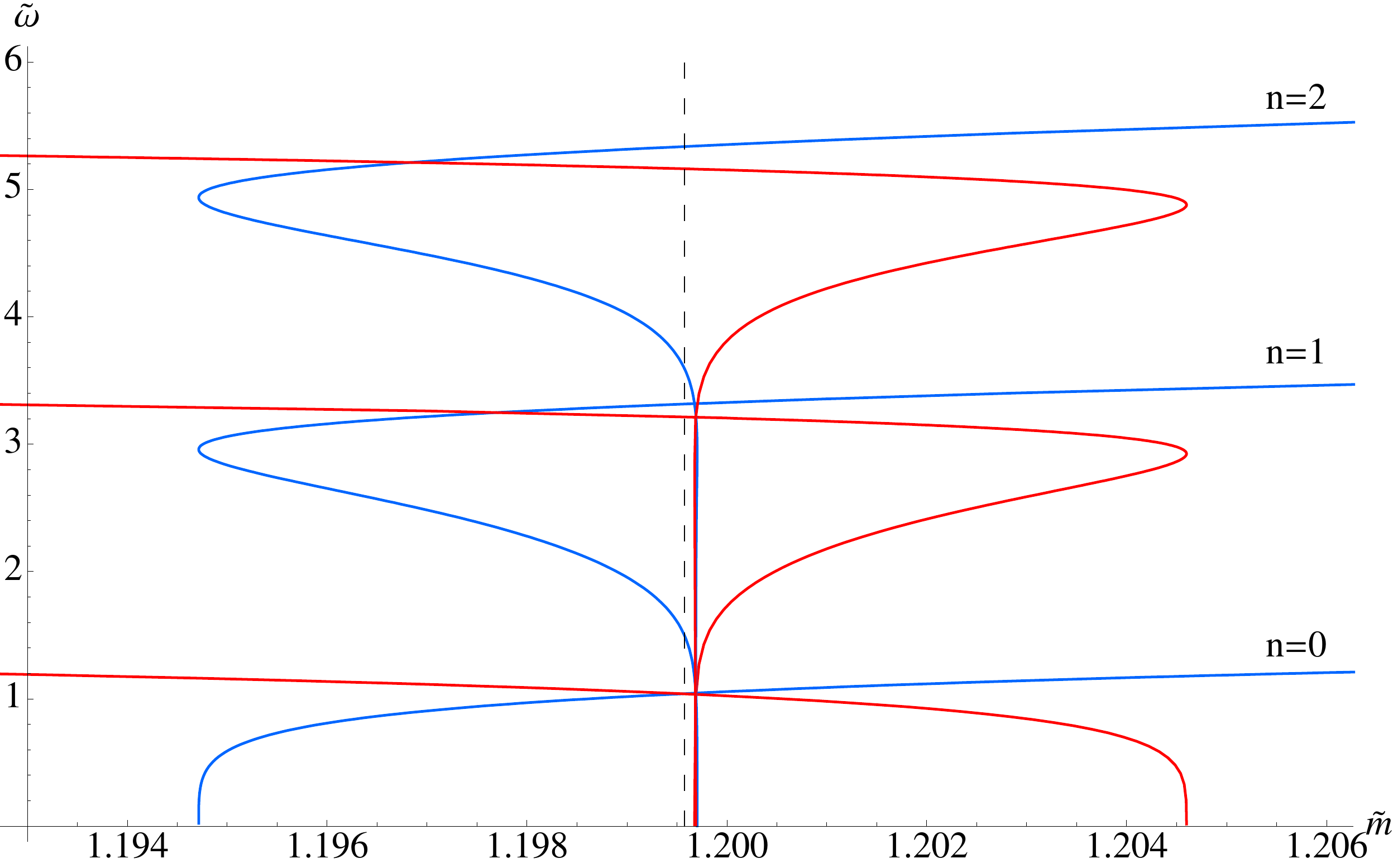} 
\caption{\small A plot of the meson mass $\tilde\omega$ versus the bare quark
  mass $\tilde m$ for the spectrum of D5 brane fluctuations corresponding to $l(r)$. At large $\tilde m$ the spectrum is described by equation (\ref{flatD3_D5}). At zero bare mass ($\tilde m=0$) the spectrum is equidistant and is described by equation (\ref{SpectSug}) with $\Delta=2$. Near the phase transition the spectrum becomes tachyonic at the points where the slope of the $-\tilde c$ versus $\tilde m$ plot from figure \ref{fig:condD5} diverges.}
\label{fig:SpectrD5}
}
Similarly to the D7--brane case, the spectrum for large $\tilde m$ matches the spectrum of the D3/D5 intersection studied in refs.~\cite{{Arean:2006pk},{Myers:2006qr}} where the following analytic expression has been obtained:
\begin{equation}
\omega=\frac{2m}{R^2}\sqrt{(n+1/2)(n+3/2)}\ .\label{flatD3_D5}
\end{equation}
For $n=0,1,2$, (\ref{flatD3_D5})  is represented by the dashed black
lines in figure \ref{fig:SpectrD5}. Interestingly at vanishing bare
mass ($\tilde m=0$), the spectrum is discrete and equidistant and is
described by equation (\ref{SpectSug}) with $\Delta=2$. This is to be
expected since fluctuations along $l$ correspond to fluctuations of
the fundamental condensate of the dual gauge theory and for the
defect--field theory that we consider (2+1 dimensional defect), the
condensate operator is a dimension two operator ($\Delta=2$). We refer
the reader to sections 4 and 5 below 
for an analytic derivation of the spectrum at zero bare mass both on
the gravity and on the field theory side of the correspondence.

The second plot in figure \ref{fig:SpectrD5} represents the structure
of the spectrum near the phase transition. From the first order phase
transition pattern of the plot of $-\tilde c$ versus $\tilde m$
presented in figure \ref{fig:condD5}, one may expect that the theory
becomes unstable at the points where the slope of the fundamental
condensate diverges. As we can see from the graph for the ground state 
($n=0$) in figure \ref{fig:SpectrD5}, the spectrum drops to zero exactly at these points. We have verified numerically that beyond these points the spectrum of the ground state is indeed tachyonic.

For completeness of our study and to verify the stability of the D5--brane
embeddings, we consider the spectrum of fluctuations along the
Neumann--Dirichlet direction~$\alpha$ of the D5--brane as defined in
\eqref{parmD5} . Note that the 
term `Neumann-Dirichlet'   is valid only at large $\tilde m$, where we
can think of the $AdS_5\times S^5$ geometry as corresponding to the
near-horizon limit of the gravitational background of a stack of coincident
D3--branes since large $\tilde m$ corresponds to a large $\tilde S^3$
radius. 
To obtain the spectrum, we expand
$\alpha=0+(2\pi\alpha')\delta\alpha$ in the DBI Lagrangian of the D5--brane
and obtain the corresponding quadratic Lagrangian. Here~$\alpha$ is the polar
angle on $\tilde S^3\subset AdS_5$ \eqref{parmD5}, with $\tilde S^3$
denoting the 3--sphere in the field theory directions. 
In general, fluctuations along~$\alpha$ couple to fluctuations of the $U(1)$ gauge field of the probe
D5--brane (as discussed in detail in
refs.~\cite{{Arean:2006pk},{Myers:2006qr}} for the ``flat" D3/D5
case). However, if one restricts the fluctuations to depend only on the time
$t$ and the holographic direction $r$, the modes decouple. It is the stability
of this decoupled mode that we analyze. Another important point is that the
corresponding gauge invariant operator is a dimension four ($\Delta=4$)
operator. This is obtained from  the asymptotic behaviour of the general
solution of the equation of motion, using the standard field/operator
dictionary, in analogy to the calculation we performed for the D7 brane case
above.

After implementing the ansatz $\delta\alpha=e^{i\omega t}h(r)$, solving numerically the corresponding equation of motion and imposing regularity of the solution, we obtain the plot of the spectrum versus bare mass presented in figure \ref{fig:SpectND}.
\FIGURE{
\centering
\includegraphics[width=4in]{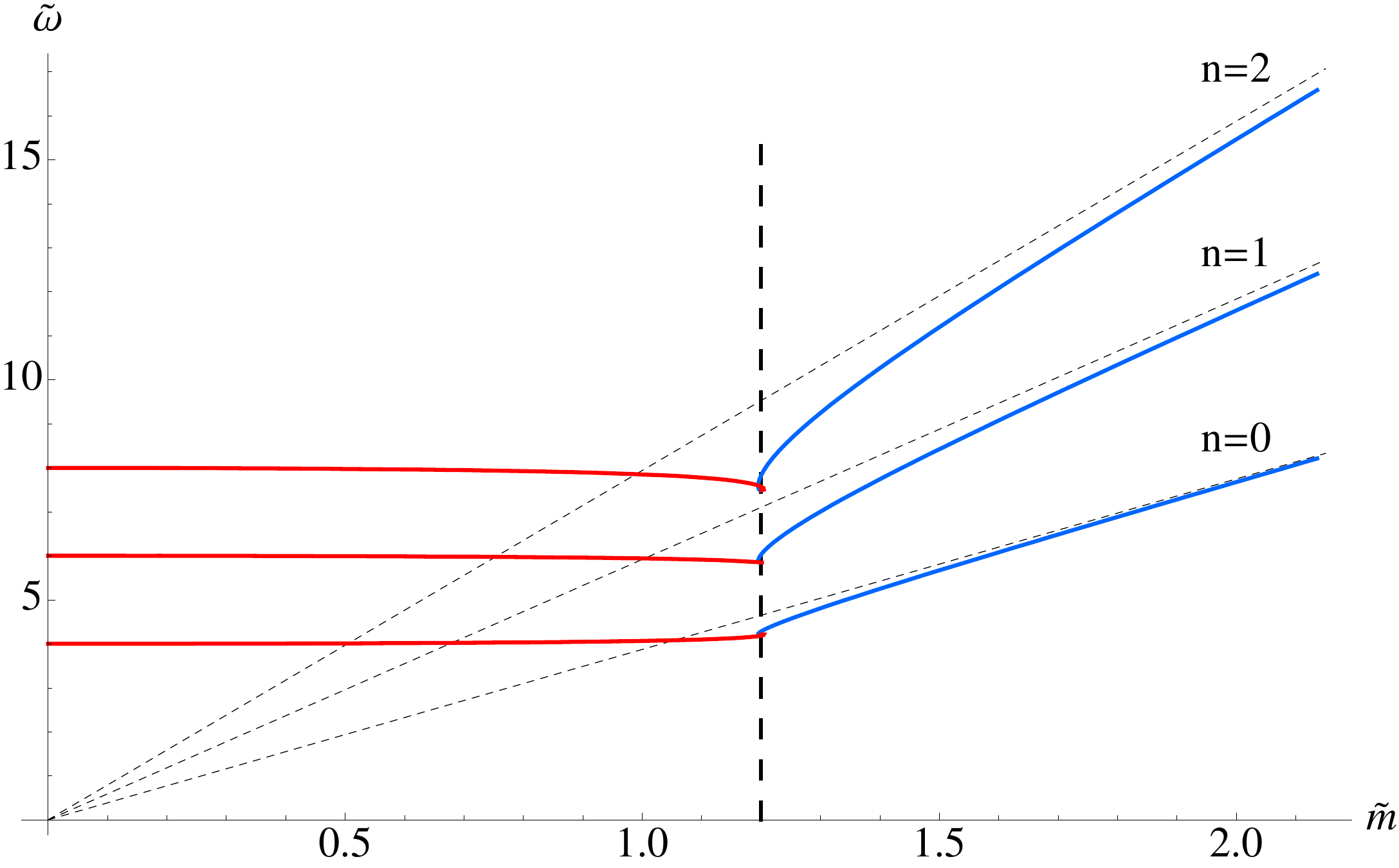} 
\includegraphics[width=2.6in]{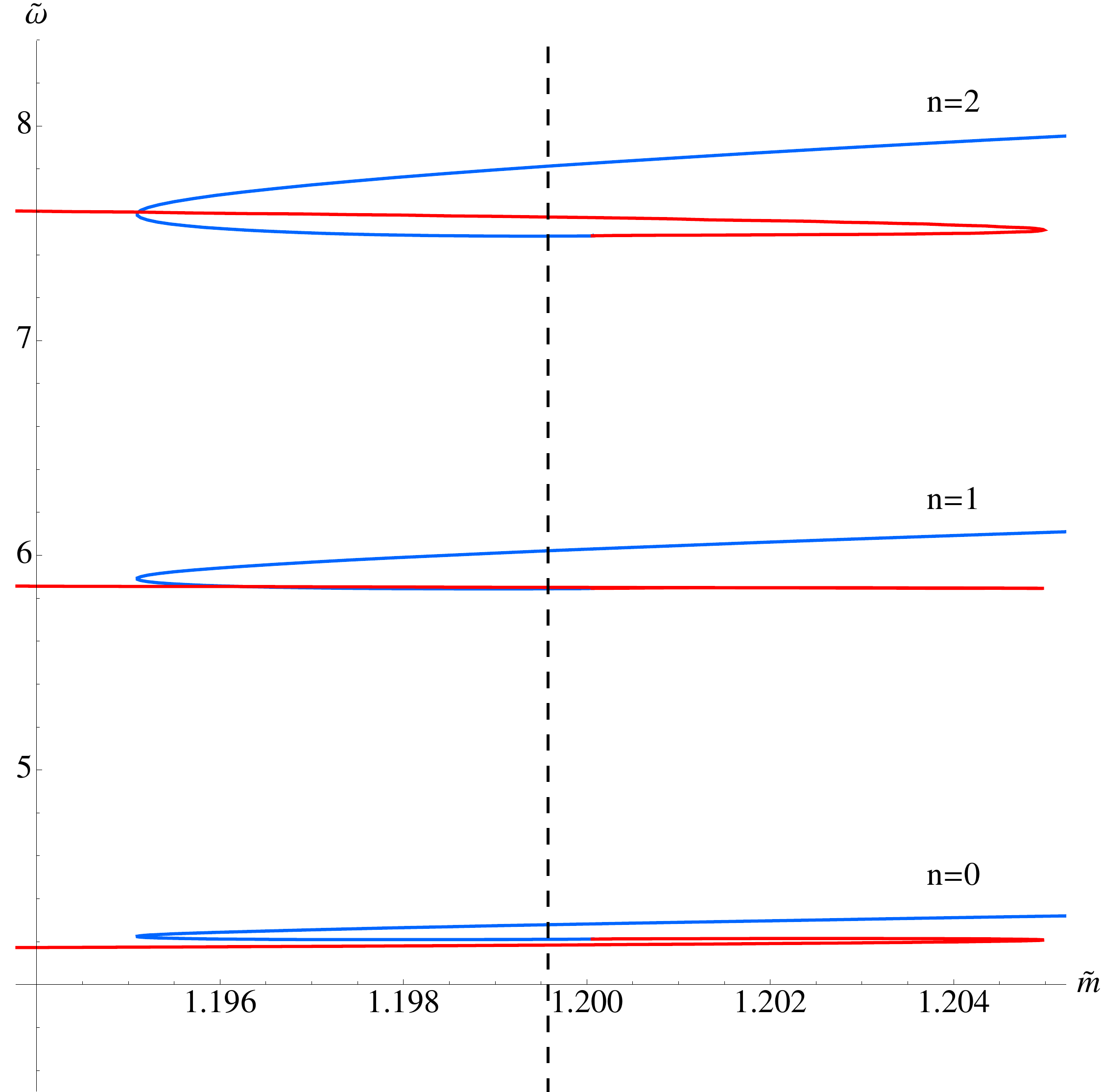} 
\caption{\small A plot of the spectrum 
$\tilde\omega$ versus the bare quark mass 
$\tilde m$ for the spectrum of fluctuations corresponding to $\alpha$. At large $\tilde m$ the spectrum is described by equation (\ref{flatD3_D5ND}), while at zero bare mass ($\tilde m=0$) the spectrum is equidistant and is described by equation (\ref{SpectSug}) with $\Delta=4$. The spectrum is tachyon free and has a finite jump at the phase transition.}
\label{fig:SpectND}
}
 
 At large $\tilde m$, the spectrum should match the spectrum of the
 ``flat'' D3/D5 case. 
For the decoupled mode subject to our investigation, the analytic result obtained in refs.~\cite{{Arean:2006pk},{Myers:2006qr}} is given by:
\begin{equation}
\omega=\frac{2m}{R^2}\sqrt{(n+3/2)(n+5/2)} \, , \label{flatD3_D5ND}
\end{equation}
represented by the dashed black lines in figure \ref{fig:SpectND}. One can see
the good agreement at large $\tilde  m$. Near the phase transition, the
spectrum remains tachyon-free. At the phase transition it has a finite jump,
as shown in the second plot in figure
\ref{fig:SpectND}. Interestingly, at zero bare mass ($\tilde m=0$) the
spectrum is again described by equation (\ref{SpectSug}) with
$\Delta=4$. Once again the energy of the ground state is determined by
its engineering dimension. A detailed analytic study of the spectrum
at zero bare quark mass is provided in  sections 4 and 5. 
The lack of tachyon modes in the spectrum suggests that the D5--brane embedding is stable and does not slip off the equatorial two sphere of the three sphere where the field theory is defined ($\alpha=0$,  $\tilde S^2\subset \tilde S^3$).

\section{Spectrum at zero bare mass -- Gravity side }
\label{sec:gravity}

Here  we provide an analytic derivation of the spectrum of
fluctuations at zero bare mass on the gravity side of the correspondence. 
To obtain the spectrum we recast the second order linear differential
equation of motion of the fluctuation modes to one dimensional
Schr\"odinger equation. Our analysis is along the lines of the one employed in ref.~\cite{Hoyos:2006gb} for the study of quasi-normal modes in the meson melting phase transition in flavoured supersymmetric Yang-Mills plasma. Let us briefly describe the method.

We start with a second order differential equation of the form:
\begin{equation}
h''(\tilde\rho)+C_1(\tilde\rho)h'(\rho)+(B(\tilde\rho)^2\tilde\omega^2+C_0(\tilde\rho))h(\tilde\rho)=0\label{A1.1}
\, .
\end{equation}
Upon the substitution $h(\tilde\rho)=\sigma(\tilde\rho)f(\tilde\rho)$, with $\sigma(\tilde\rho)$ defined via
\begin{equation}
\frac{\sigma'(\tilde\rho)}{\sigma(\tilde\rho)}=-\frac{1}{2}\left(C_1(\tilde\rho)+\frac{B'(\tilde\rho)}{B(\tilde\rho)}\right)
\, , \end{equation}
equation (\ref{A1.1}) becomes:
\begin{equation}
\frac{1}{B(\tilde\rho)}\frac{d}{d\tilde\rho}\left(\frac{1}{B(\tilde\rho)}\frac{d}{d\tilde\rho}f(\tilde\rho)\right)+\left(\tilde\omega^2-V_{\rm{eff}}(\tilde\rho)\right)f(\tilde\rho)=0\, ,
\end{equation}
where the effective potential $V_{\rm eff}(\tilde\rho)$ is given by
\begin{equation}
V_{\rm{eff}}(\tilde\rho)=-\frac{1}{B(\tilde\rho)^2}\left(C_0(\tilde\rho)+C_1(\tilde\rho)\frac{\sigma'(\tilde\rho)}{\sigma(\tilde\rho)}+\frac{\sigma''(\tilde\rho)}{\sigma(\tilde\rho)} \right)\ .\label{effpotRho}
\end{equation}
The final step is to define a new variable $u(\tilde\rho)$ satisfying ${dz}/{d\tilde\rho}=B(\tilde\rho)$. The resulting equation of motion is equivalent to a one dimensional Schr\"odinger equation for a particle with $2mE=\tilde\omega^2$ in the effective potential $V_{\rm eff}(z)$:
\begin{equation}
f''(z)+[\tilde\omega^2-V_{\rm eff}(z)]f(z)=0\ .\label{Schrod}
\end{equation}
This approach is particularly useful since it enables us to study
properties of the spectrum, such as the existence of bound states,
even without solving the equation of motion. In our case the
corresponding effective potential is relatively simple and we are able
to obtain the spectrum in closed form. Let us present our results for
the different modes studied in section \ref{sec:mesonspectra}. 
\subsection{Fluctuations of a D7--brane probe}
\subsubsection{Fluctuations along $L$}
We consider the ansatz $\delta L=e^{i\frac{\tilde\omega}{R} t}h(\tilde\rho)\tilde{\cal Y}^{\tilde l}({\tilde S^3}){\cal Y}^{l}({S^3})$ in the equation of motion derived from the quadratic lagrangian (\ref{LagrFlD7}). Here $\tilde\rho=\rho/R$, while $\tilde{\cal Y}^{\tilde l}$ and ${\cal Y}^l$ are spherical harmonics on $\tilde S^3\subset$ AdS$_5$ and $S^3\subset S^5$ respectively. The corresponding differential equation for $h(\tilde\rho)$ is of the general form (\ref{A1.1}) with coefficients:
\begin{gather}
C_0(\tilde\rho)=-\frac{\tilde l(\tilde
  l+2)}{(\tilde\rho^2-\frac{1}{4})^2}-\frac{l(l+2)}{\tilde\rho^2}+\frac{8+16\tilde\rho^2}{\tilde\rho^2-16\tilde\rho^6};
\nonumber\\ C_1(\tilde\rho)=\frac{5+16\tilde\rho
  ^2+48 \tilde\rho ^4}{-\tilde\rho +16 \tilde\rho
  ^5};~~B(\tilde\rho)=\frac{4}{1+4\tilde\rho^2} \, .\label{CoeffD7L}
\end{gather}
Applying the prescription outlined in equations (\ref{A1.1})-(\ref{Schrod}) results in the effective potential
\begin{equation}
V_{\rm eff}(z)=\frac{3+4l(l+2)}{4\sin^2z}+\frac{3+4\tilde l(\tilde l+2)}{4\cos^2 z};~~~~~~~z\in\left[{\pi}/{2},\pi\right); \ ,\label{effectPotnL}
\end{equation}
where 
\begin{equation}
z(\tilde\rho)\equiv 2\arctan{(2\tilde\rho)}. \label{ZofRho}
\end{equation}
The effective potential diverges at the interval boundaries $z=\pi/2,
\pi$,  implying that the corresponding ``wave function" $f(z)$ should
vanish there. It turns out that for this effective potential, we can solve equation (\ref{Schrod}) exactly. 
The solution regular at $z=\pi/2$ is given by:
\begin{equation}
f^{\pm}(z)=C(\cos z)^{\tilde
  l+\frac{3}{2}}{}_2F_1[\frac{1}{2}(3+\tilde
l+l-\tilde\omega),\frac{1}{2}(3+\tilde l+l+\tilde\omega),\tilde
l+2,\cos^2 z](\sin z)^{l+\frac{3}{2}} \, .
\end{equation}
Regularity at $z=\pi$ requires that one of the first two arguments of
the hypergeometric function 
is a non-positive integer. Without loss of generality, we consider
only positive $\tilde\omega$. This implies that
\begin{equation}
\frac{1}{2}(3+\tilde l+l-\tilde\omega)=-n;~~~~~~~~~n=0,1,2,\dots\ .
\end{equation}
Therefore we obtain the following expression for the spectrum:
\begin{equation}
\tilde\omega=3+l+2n+\tilde l;~~~~~~~~~n=0,1,2,\dots\ .\label{SpectrumD7m0}
\end{equation}
At $l=\tilde l=0$, (\ref{SpectrumD7m0}) 
agrees with the numerical analysis considered in section
\ref{sec:mesonspectra} above. 
1
\subsubsection{Fluctuations along $\phi$.}
In this subsection we study the spectrum of fluctuations along $\phi$
in the limit of zero bare quark mass. 
Our starting point is equation (\ref{EOMPhi}). In this case there is a
subtlety in taking the limit $\bar L\to 0$ in equation
(\ref{EOMPhi}). Indeed if we substitute $\bar
L(\rho)=\epsilon\xi(\rho)$ into equation (\ref{EOMPhi}) and take the
limit $\epsilon\to 0$, it is easy to verify that the coefficients of
the equation of motion will remain $\xi(\rho)$ dependent. Indeed  if
we bring the equation of motion to the general form (\ref{A1.1}), with
dimensionless variables $\tilde\omega=\omega R$ and
$\tilde\rho=\rho/R$),  the coefficients are:
\begin{gather}
C_0(\tilde\rho)= -\frac{\tilde l (\tilde l+2)}{(\tilde\rho
  ^2-\frac{1}{4})^2}-\frac{l (l+2)}{\tilde\rho
  ^2}; \nonumber\\ C_1(\tilde\rho)=-\frac{5+16 \tilde\rho ^2+48\tilde\rho
  ^4}{\tilde\rho -16 \tilde\rho ^5}+\frac{2 \xi '(\tilde\rho )}{\xi
  (\tilde\rho )};~~~B(\tilde\rho)=\frac{4}{1+4 \tilde\rho ^2} \, .
\end{gather}
The resulting effective potential $V_{\rm eff}(\tilde\rho)$ defined
via equation (\ref{effpotRho}) depends on the function
$\xi(\tilde\rho)$ and its derivatives. However $\xi(\tilde\rho)$
should also satisfy the linearized equation of motion for the
classical D7--brane embedding. The next step is thus to solve for the highest derivative of $\xi(\tilde\rho)$ from the linearized classical equation of motion and substitute into the expression for the effective potential $V_{\rm{eff}}(\tilde\rho)$. It is a straightforward exercise to verify that this completely removes the $\xi(\tilde\rho)$ dependence of the effective potential. 

Note that the function $B(\tilde\rho)$ is the same as in equation
(\ref{CoeffD7L}) and hence $z(\tilde\rho)$ is given by equation
(\ref{ZofRho}). In this coordinate $z$,  the equation of motion is of the general form (\ref{Schrod}) with $V_{\rm  eff}(z)$ given by equation (\ref{effectPotnL}). Therefore we conclude that the spectrum of fluctuations along $\phi$ at zero bare mass is identical to the spectrum of fluctuations along $L$ and is given by equation (\ref{SpectrumD7m0}) which we duplicate below:
\begin{equation}
\tilde\omega=3+l+2n+\tilde l;~~~~~~~~~n=0,1,2,\dots\ .\label{SpectrumD7m0phi}
\end{equation}
One can see that at $\tilde l=0$ equation (\ref{SpectrumD7m0phi})
agrees with the numerical results for the spectrum of fluctuations
along $\phi$ from Figure 5 in Section 3. 

\subsubsection{Fluctuations of the gauge field.}
In this subsection we are interested in studying the fluctuations of the $U(1)$ gauge field of the probe D7--brane at zero bare mass. Our starting point is the action of the probe D7--brane:
\begin{equation}
S=\frac{\mu_7}{g_s}\int d\xi^8\sqrt{|G_{ab}+(2\pi\alpha')^2F_{ab}|}+\frac{(2\pi\alpha')^2}{2}\mu_7\int P[C_{(4)}]\wedge F_{(2)}\wedge F_{(2)}\ ,\label{actionGF}
\end{equation}
where $P[C_4]$ is the pullback of the R--R four form whose ``electric part'' is defined by
\begin{equation}
dC_{(4)}=\frac{4}{R}{\rm Vol}({\rm AdS}_5)\ ,
\end{equation}
where ${\rm Vol}({\rm AdS}_5)$ is the volume form of AdS$_5$. Note that in (\ref{actionGF}) we have written only the quadratic contribution (in $\alpha'$) to the Wess--Zumino term of the action. The equation of motion is:
\begin{equation}
\sqrt{-g}\nabla_a F^{ab}-\frac{4}{R}\rho u^2\left(1+\frac{R^2}{4u^2}\right)\left(1-\frac{R^2}{4u^2}\right)^3\left(1+\frac{L}{\rho}L'\right)\sqrt{\tilde g_3}\delta_k^b{\tilde\varepsilon}^{kij}\partial_i A_j=0\ ,\label{EOMGAUGE}
\end{equation}
where $a,b$ are general indices for the eight worldvolume coordinates
of the probe D7--brane, while $i,j,k$ are indices for the coordinates
parametrizing the $S^3\subset S^5$ wrapped by the D7--brane. Note
also that the Levi-Civita symbol $\tilde\varepsilon^{ijk}$ is a tensor
density (takes values $0,\pm 1$) and $\tilde g_3$ is the determinant
of the metric on the unit $\tilde S^3\subset $ AdS$_5$. In this
section we are interested in the spectrum of fluctuations at zero bare
quark mass ($L(\rho)\equiv 0$). Thus we write
\begin{equation}
\sqrt{-g}=R^6{\cal G}({\tilde \rho})\sqrt{\tilde
  g_3}\sqrt{g_3};~~~~~~~~{\cal G}({\tilde
  \rho})=\tilde\rho^3\left(1+\frac{1}{4\tilde\rho^2}\right)\left(1-\frac{1}{4\tilde\rho^2}\right)^3;~~\tilde\rho=\frac{\rho}{R}
\ .
\end{equation}
Here ${g_3}$ is the determinant of the metric on the unit $S^3\subset S^5$. With these notations the equation of motion (\ref{EOMGAUGE}) can be written as
\begin{equation}
\partial_a({\cal G}({\tilde \rho})\sqrt{\tilde
  g_3}\sqrt{g_3}F^{ab})-\frac{4}{R^4}{\cal G}({\tilde
  \rho})\sqrt{\tilde g_3}\delta_k^b{\tilde\varepsilon}^{kij}\partial_i
A_j=0\, .\label{EOMGAUGE1}
\end{equation}
Since we are in global coordinates, our probe brane is wrapping two
three--spheres: 
one in the AdS$_5$ part of the geometry, $\tilde S^3\subset$ AdS$_5$, 
and one in the $S^5$ part of the geometry, $S^3\subset S^5$. Moreover, we
have to expand both in scalar and in vector 
spherical harmonics on the $\tilde S^3$. Let us briefly review some basic
properties of spherical harmonics:

The scalar spherical harmonics ${\cal Y}^l(\Omega_3)$ on $S^3$ satisfy:
\begin{equation}
\nabla_i\nabla^i {\cal Y}^l=-l(l+2){\cal Y}^l\ ,
\end{equation}
and form a $(\frac{l}{2},\frac{l}{2})$ irreducible representation of
$SO(4)$. On the other hand,
there are three types of vector spherical harmonics on $S^3$ \cite{Kruczenski:2003be}: The first type is constructed by simply applying a derivative on the scalar spherical harmonics $\partial_i {\cal Y}^l$ (longitudinal vector spherical harmonics) and transform in the $(\frac{l}{2},\frac{l}{2})$ irreducible representation of $SO(4)$. The other two types ${\cal Y}^{l,\pm}_i(\Omega_3)$ (transverse vector spherical harmonics) transform in the $(\frac{l\mp1}{2},\frac{l\pm1}{2})$ irreducible representations and satisfy: 
\begin{eqnarray}
\nabla_i\nabla^i{\cal Y}^{l,\pm}_j-{\cal R}_j^k{\cal Y}^{l,\pm}_k&=&-(l+1)^2{\cal Y}^{l,\pm}_j\ ,\\
\varepsilon_i^{~jk}\nabla_j{\cal Y}^{l,\pm}_k&=&\pm (l+1){\cal Y}^{l,\pm}_i \ , \\
\nabla^i{\cal Y}^{l,\pm}_i&=&0\ .
\end{eqnarray}
Here ${\cal R}_i^j=2\delta_i^j$ is the Ricci scalar curvature of an unit $S^3$.
Next we classify the fluctuations of the $U(1)$ gauge field into three different types:

{\bf Type I} comprises of fluctuations $A^{\pm}_i$ along the $S^3
\subset S^5$ which have expansion  in transverse vector spherical
harmonics ${\cal Y}^{l,\pm}_i$. Since these modes satisfy $\nabla^i
A^{\pm}_i=0$, they decouple from the remaining modes. Hence we have the ansatz:  
\begin{equation}
A_i^{\pm}=\Phi_{\rm I}^{\pm}(\rho)e^{i \omega t}\tilde{\cal Y}^{\tilde
  l}(\tilde S^3){\cal
  Y}^{l,\pm}_i(S^3);~~~A_0=0;~~~A_{\alpha}=0;~~~A_{\rho}=0 \, . \label{ANZI}
\end{equation}
Here greek indices denote components along the $\tilde S^3\subset$ AdS$_5$.

{\bf Type II} are fluctuations of the gauge field $A_{\alpha}^{\pm}$
along the $\tilde S^3\subset$ AdS$_5$, 
 which have an expansion in transverse vector spherical harmonics $\tilde{\cal Y}^{l,\pm}_i$. Similarly to type I modes, these modes satisfy $\nabla^{\alpha } A_{\alpha}=0$ and decouple from the rest. This implies the ansatz:
\begin{equation} 
A_{\alpha}^{\pm}=\Phi_{\rm II}^{\pm}(\rho)e^{i \omega t}\tilde{\cal
  Y}^{\tilde l,\pm}_{\alpha}(\tilde S^3){\cal
  Y}^l(S^3);~~~A_0=0;~~~A_{\rho}=0;~~~A_{i}=0 \label{ANZII}\, .
\end{equation}

Finally {\bf type III} modes are modes that do not fall into type I or
type II classes. One may expect that in this case all modes are
coupled and should be studied simultaneously. However only vector
spherical harmonics transforming as scalar spherical harmonics can be
used, which  are given by $\partial_{\alpha}\tilde{\cal Y}^{\tilde
  l}(\tilde S^3)$ and  by $\partial_{i}{\cal Y}^{l}(S^3)$. Therefore
both $A_{\alpha}$ and $A_i$ components of the gauge field are pure
gauge and one of these sets may always be gauged away. 
Thus without loss of generality, the type III modes can be described by the ansatz:
\begin{eqnarray}
A_0&=&\hat\Phi_{\rm III}(\rho)e^{i\omega t}\tilde{\cal Y}^{\tilde l}(\tilde S^3){\cal Y}^{l}(S^3);~~A_{\rho}=\Phi_{\rm III}(\rho)\tilde{\cal Y}^{\tilde l}(\tilde S^3){\cal Y}^{l}(S^3);\label{ANZIII}\\
A_{i}&=&\tilde\Phi_{\rm III}(\rho)\tilde{\cal Y}^{\tilde l}(\tilde
S^3)\partial_i {\cal Y}^{l}(S^3);~~A_{\alpha}=0 \, . \nonumber
\end{eqnarray}

Next we proceed with the analysis of the equations of motion. Let us begin with {\bf type I} modes. After substituting the ansatz (\ref{ANZI}) into the equations of motion (\ref{EOMGAUGE1}) we obtain:
\begin{eqnarray}
\frac{1}{{\cal
    G(\tilde\rho)}\tilde\rho^2}\partial_{\tilde\rho}\left({\cal
    G}(\tilde\rho)\tilde\rho^2\partial_{\tilde\rho}\Phi^{\pm}_{\rm
    I}\right)+\left[\frac{\tilde\omega^2}{(\tilde\rho^2+\frac{1}{4})^2}-\frac{\tilde
    l(\tilde
    l+2)}{(\tilde\rho^2-\frac{1}{4})^2}-\frac{(l+1)(l+1\pm4)}{\tilde\rho^2}\right]\Phi^{\pm}_{\rm
  I}=0 \, . \nonumber\\ \label{EOMAPM}
\end{eqnarray}
Equation (\ref{EOMAPM}) is of the general form (\ref{A1.1}) with coefficients given by:
\begin{equation}
C_1({\tilde\rho})=\frac{3+16\tilde\rho^2+80\tilde\rho^4}{\tilde\rho(16\tilde\rho^4-1)};~~B(\tilde\rho)=\frac{4}{4\tilde\rho^2+1}~~C_0(\tilde\rho)=-\frac{\tilde
  l(\tilde
  l+2)}{(\tilde\rho^2-\frac{1}{4})^2}-\frac{(l+1)(l+1\pm4)}{\tilde\rho^2}
\, . 
\end{equation}
Using the procedure outlined in equations (\ref{A1.1})-(\ref{Schrod}), we obtain an equation of the form (\ref{Schrod}) with effective potential
\begin{equation}
V^{\pm}_{\rm eff}(z)=\frac{(2l+1\pm4)(2l+3\pm4)}{4\sin^2z}+\frac{3+4\tilde l(\tilde l+2)}{4\cos^2 z};~~~z \in \left[\pi/2,\pi\right)\ ,
\end{equation}
where $z(\tilde\rho)$ is given by equation (\ref{ZofRho}). The general solution regular at $z=\pi/2$ is given by:
\begin{gather}
f^{\pm}(z)=C_{\pm}(\cos z)^{\tilde l+\frac{3}{2}} \hspace{12cm} 
\nonumber\\ \hspace{1.6cm} \times \; {}_2F_1[\frac{1}{2}(3\pm2+\tilde
l+l-\tilde\omega),\frac{1}{2}(3\pm2+\tilde l+l+\tilde\omega),\tilde
l+2,\cos^2 z](\sin z)^{l+\frac{3}{2}\pm2} \, .
\end{gather}
Regularity at $z=\pi$ requires that one of the first two arguments of
the hypergeometric function must be  a non-positive integer. Without
loss of generality we can consider only positive $\tilde\omega$ and
find 
\begin{equation}
\frac{1}{2}(3\pm2+\tilde l+l-\tilde\omega)=-n;~~~~~~~~~n=0,1,2,\dots\ .
\end{equation}
Therefore we obtain the following expression for the spectrum:
\begin{equation}
\tilde\omega^{\pm}=3\pm2+l+2n+\tilde l;~~~~~~~~~n=0,1,2,\dots\ .
\end{equation}
At large $\tilde\rho$ the equations of motion (\ref{EOMAPM}) are the same as the equations of motion of type I modes for the ``flat case" studied in ref.~\cite{Kruczenski:2003be}. Therefore the conformal dimension of the corresponding gauge invariant operators are the same, namely $\Delta_{\pm}=3\pm 2+l$. Finally for the spectrum of type I fluctuations we can write:
\begin{equation}
\tilde\omega^{\pm}=\Delta_{\pm}+2n+\tilde l~~~~~~~~~n=0,1,2,\dots\ .\label{SPECTRUMPM}
\end{equation}
We see that similarly to the spectrum of fluctuations of the transverse scalars, the ground state is again determined by the conformal dimension of the corresponding gauge invariant operators. In the next section we will confirm equation (\ref{SPECTRUMPM}) from field theory considerations.

We now focus on the spectrum of {\bf type II} modes. Substitution of the ansatz (\ref{ANZII}) into the equations of motion (\ref{EOMGAUGE1}) results in the following equation:
\begin{equation}
\frac{16\tilde\rho}{(16\tilde\rho^4-1)}\partial_{\tilde\rho}\left(\frac{(16\tilde\rho^4-1)}{16\tilde\rho}\partial_{\tilde\rho}\Phi_{\rm II}^{\pm}\right)+\left[\frac{\tilde\omega^2}{(\tilde\rho^2+\frac{1}{4})^2}-\frac{(\tilde l+1)^2}{(\tilde\rho^2-\frac{1}{4})^2}-\frac{l(l+2)}{\tilde\rho^2}\right]\Phi_{\rm II}^{\pm}=0\ ,
\end{equation}
which is of the type (\ref{A1.1}) with coefficients:
\begin{gather}
C_1(\tilde\rho)=-\frac{1+48\tilde\rho^4}{\tilde\rho-16\tilde\rho^5};~~~B(\tilde\rho)=\frac{4}{4\tilde\rho^2+1};
\nonumber\\ C_0(\tilde\rho)=-\frac{(\tilde l+1)^2}{(4\tilde\rho^2-1)^2}-\frac{l(l+2)}{\tilde\rho^2} \, .
\end{gather}
Following the procedure outlined in (\ref{A1.1})-(\ref{Schrod}) one obtains an equation of the form (\ref{Schrod}) with effective potential
\begin{equation}
V_{\rm eff}(z)=\frac{3+4\tilde l(\tilde l+2)}{4\cos^2 z}+\frac{3+4l(l+2)}{4\sin^2 z} \, ,
\end{equation}
where $z(\tilde\rho)$ is given by equation (\ref{ZofRho}). The general solution regular at $z=\pi/2$ is given by:
\begin{equation}
f[z]=C(\cos z)^{\frac{3}{2}+\tilde l}{}_2F_1[\frac{1}{2}(3+\tilde l+l-\tilde\omega),\frac{1}{2}(3+\tilde l+l+\tilde\omega),\tilde l+2,\cos^2 z](\sin z)^{\frac{3}{2}+l}\ .
\end{equation}
Regularity at $z=\pi$ requires that one of the first two arguments of the hypergeometric function be a non-positive integer. Therefore the spectrum of type II modes is given by:
\begin{equation}
\tilde\omega=3+l+2n+\tilde l;~~~~~~~~~n=0,1,2,\dots\ .
\end{equation}
Given that the conformal dimension of the dual gauge invariant operator is $\Delta=l+3$ \cite{Kruczenski:2003be} one obtains:
\begin{equation}
\tilde \omega=\Delta+2n+\tilde l;~~~~~~~~~n=0,1,2,\dots\ .
\end{equation}
Once again the ground state is determined by the conformal dimension of the dual gauge invariant operator.

Finally we study the spectrum of {\bf type III} modes. Substitution of the ansatz (\ref{ANZIII}) into (\ref{EOMGAUGE1}) results in the following equations coming from components of the gauge field along the $t,\rho$, $S^3\subset S^5$ and $\tilde S^3\subset$ AdS$_5$ directions:
\begin{eqnarray}
&&\frac{(\tilde\rho^2+\frac{1}{4})^2}{\tilde\rho^4{\cal G}(\tilde\rho)}\partial_\rho(\frac{\tilde\rho^4{\cal G}(\tilde\rho)}{(\tilde\rho^2+\frac{1}{4})^2}(\partial_{\tilde\rho}\hat\Phi_{\rm III}-i\tilde\omega\Phi_{\rm III}))-\frac{\tilde l(\tilde l+2)}{(\tilde\rho^2-\frac{1}{4})^2}\hat\Phi_{\rm III}-\frac{l(l+2)}{\tilde\rho^2}(\hat\Phi_{\rm III}-i\omega\tilde\Phi_{\rm III})=0\nonumber\ ,\\
&&\frac{1}{(\tilde\rho^2+\frac{1}{4})^2}(\tilde\omega^2\Phi_{\rm III}+i\tilde\omega\partial_{\tilde\rho}\hat\Phi_{\rm III})-\frac{\tilde l(\tilde l+2)}{(\tilde\rho^2-\frac{1}{4})^2}\Phi_{\rm III}-\frac{l(l+2)}{\tilde\rho^2}(\Phi_{\rm III}-\partial_{\tilde\rho}\tilde\Phi_{\rm III})=0\nonumber\ ,\\
&&\frac{1}{\tilde\rho^2{\cal G}(\tilde\rho)}\partial_{\tilde\rho}(\tilde\rho^2{\cal G}(\tilde\rho)(\partial_{\tilde\rho}\tilde\Phi_{\rm III}-\Phi_{\rm III}))+\frac{1}{(\tilde\rho^2+\frac{1}{4})^2}(\tilde\omega^2\tilde\Phi_{\rm III}+i\tilde\omega\hat\Phi_{\rm III})-\frac{\tilde l(\tilde l+2)}{(\tilde\rho^2-\frac{1}{4})^2}\tilde\Phi_{\rm III}=0\nonumber\ , \\
&&\frac{(\tilde\rho^2-\frac{1}{4})^2}{\tilde\rho^4{\cal G}(\tilde\rho)}\partial_\rho(\frac{\tilde\rho^4{\cal G}(\tilde\rho)}{(\tilde\rho^2-\frac{1}{4})^2}\Phi_{\rm III})-\frac{i\tilde\omega}{(\tilde\rho^2+\frac{1}{4})^2}\hat\Phi_{\rm III}-\frac{l(l+2)}{\tilde\rho^2}\tilde\Phi_{\rm III}=0 \ . \label{EOMIII}
\end{eqnarray} 
Note that the third and fourth equations in (\ref{EOMIII}) are valid
only for $l>0$ and $\tilde l>0$, respectively. Moreover, note that we
have four equations for three unknown functions $\hat\Phi_{\rm III}$,
$\Phi_{\rm III}$ and $\tilde\Phi_{\rm III}$. However one can show that
only three of the equations are independent. In particular, it is easy
to verify with simple algebraic manipulations that the last three
equations in (\ref{EOMIII}) imply the first one, which may thus be
skipped. 
Moreover, generically $\tilde\omega\neq 0$ and one can use the fourth equation to eliminate $\hat\Phi_{\rm III}$ from the second and third equations in (\ref{EOMIII}). The resulting system  of coupled equations for $\Phi_{\rm III}$ and $\tilde\Phi_{\rm III}$ is given by:
\begin{eqnarray}
&&\Phi_{\rm III}''(\tilde\rho)+C_{11}(\tilde\rho)\Phi_{\rm III}'(\tilde\rho)+[B_1(\tilde\rho)^2\tilde\omega^2+C_{01}(\tilde\rho)]\Phi_{\rm III}+M_{12}(\tilde\rho)\tilde\Phi_{\rm III}(\tilde\rho)=0\ ,\nonumber\\
&&\tilde\Phi_{\rm III}''(\tilde\rho)+C_{12}(\tilde\rho)\tilde\Phi_{\rm III}'(\tilde\rho)+[B_2(\tilde\rho)^2\tilde\omega^2+C_{02}(\tilde\rho)]\tilde\Phi_{\rm III}+M_{21}(\tilde\rho)\Phi_{\rm III}(\tilde\rho)=0\ ,\label{coupledIII}
\end{eqnarray}
where the coefficient functions are given by:
\begin{eqnarray}
C_{11}(\tilde\rho)&=&-\frac{1-16 \tilde\rho ^2+112 \tilde\rho ^4}{\tilde\rho -16 \tilde\rho ^5};~~B_1(\tilde\rho)=\frac{4}{1+4\tilde\rho^2};~~M_{12}(\tilde\rho)=\frac{2l(l+2)(1-4\tilde\rho^2)}{\tilde\rho^3+4\tilde\rho^5};\nonumber\\ 
C_{01}(\tilde\rho)&=&\frac{1 + 16 \tilde\rho^2\left (-1 + 2 \tilde\rho^2 \right)\left (1 + 12 \tilde\rho^2 + 72 \tilde\rho^4 \right)}{(\tilde\rho-16\tilde\rho^5)^2}-\frac{\tilde l(\tilde l+2)}{(\tilde\rho^2-\frac{1}{4})^2}-\frac{l(l+2)}{\tilde\rho^2}; \\
C_{12}(\tilde\rho)&=&-\frac{3+16 \tilde\rho ^2+80 \tilde\rho ^4}{\tilde\rho -16 \tilde\rho ^5};~~~B_2(\tilde\rho)=\frac{4}{1+4\tilde\rho^2};~~M_{21}(\tilde\rho)=\frac{(2+8\tilde\rho^2)}{\tilde\rho-4\tilde\rho^3};\nonumber\\ 
C_{02}(\tilde\rho)&=&-\frac{\tilde l(\tilde
  l+2)}{(\tilde\rho^2-\frac{1}{4})^2}-\frac{l(l+2)}{\tilde\rho^2} \, .
\nonumber
\end{eqnarray}
Note that the functions $B_1$ and $B_2$ in equations (\ref{coupledIII}) are the same. This suggests that we may employ similar procedure to the one outlined in equations (\ref{A1.1})-(\ref{Schrod}). To this end we define:
\begin{equation}
\Phi_{\rm
  III}(\tilde\rho)=f_1(\tilde\rho)\sigma_1(\tilde\rho);~~~\tilde\Phi_{\rm
  III}(\tilde\rho)=f_2(\tilde\rho)\sigma_2(\tilde\rho) \,  ,
\end{equation}
where $\sigma_1$ and $\sigma_2$ are defined via:
\begin{equation}
\frac{\sigma_i'}{\sigma_i}=-\frac{1}{2}\left(C_{1i}+\frac{B_i'}{B_i}\right);~~~~~i=1,2 ; \ .
\end{equation}
The system (\ref{coupledIII}) may be brought to the form
\begin{eqnarray}
&f_1''(z)&+[\tilde\omega^2-V^{(1)}_{\rm eff}(z)]f_1(z)+m_{12}(z)f_2(z)=0\ ,\\
&f_2''(z)&+[\tilde\omega^2-V^{(2)}_{\rm eff}(z)]f_2(z)+m_{21}(z)f_1(z)=0;\nonumber\ 
\end{eqnarray}
where 
\begin{eqnarray}
&V^{(1)}_{\rm eff}(z)&=\frac{-1+4l(l+2)}{4\sin^2 z}+\frac{3+4\tilde l(\tilde l+2)}{4\cos^2 z};~~~m_{12}(z)=-\frac{2l(l+2)}{\sin^2 z}; \label{SchrodIII}\\
&V^{(2)}_{\rm eff}(z)&=\frac{15+4l(l+2)}{4\sin^2 z}+\frac{3+4\tilde l(\tilde l+2)}{4\cos^2 z};~~~m_{21}(z)=-\frac{2}{\sin^2 z}; ~~~z\in\left[\pi/2,\pi\right)\nonumber\ .
\end{eqnarray}
and $z(\tilde\rho)$ is given by equation (\ref{ZofRho}). Note that the difference of the effective potentials $\Delta V_{\rm eff}=V^{(2)}_{\rm eff}-V^{(1)}_{\rm eff}$ and the ``coupling functions" $m_{12},m_{21}$ are all proportional to $1/\sin^2 z$ and differ only by constant multiplicative factors. This fortunate property enable us to decouple easily the system of equations (\ref{SchrodIII}). Indeed let us form the linear combination $\eta=f_1+a f_2$ for some constant parameter $a$. Combining the equations in (\ref{SchrodIII}) results in the following equation of motion for $\eta(z)$:
\begin{equation}
\eta''(z)+[\tilde\omega^2-V^{(1)}_{\rm eff}]\eta(z)+am_{21}(f_1+\frac{m_{12}-a\Delta V_{\rm eff}}{am_{21}}f_2)=0 \ .
\end{equation}
Note that due to the property mentioned above, the fraction
$(m_{12}-a\Delta V_{\rm eff})/(am_{21})$ is a constant and therefore
we can demand that it is equal to $a$. We may now write
\begin{eqnarray}
\eta''(z)+[\tilde\omega^2-V_{\rm eff}(z)]\eta(z)=0; ~~~~V_{\rm eff}(z)\equiv V^{(1)}_{\rm eff}(z)-am_{21}(z)\ .
\end{eqnarray} 
The parameter $a$ is determined by:
\begin{equation}
a=\frac{(m_{12}-a\Delta V_{\rm eff})}{(am_{21})};~ \Leftrightarrow ~a^2-2a-l(l+2)=0;~~\Rightarrow~~ a_+=l+2;~a_-=-l; \ .
\end{equation}
Therefore we end up with two decoupled modes $\eta_{\pm}(z)=f_1(z)+a_{\pm}f_2(z)$ described by the effective potentials:
\begin{equation}
V_{\rm eff}^{(\pm)}(z)=\frac{-1+4(l+1\pm 1)^2}{4\sin^2 z}+\frac{3+4\tilde l(\tilde l +2)}{4\cos^2 z}\ .
\end{equation}
The corresponding solutions of the equations of motion regular at $\pi/2$ are given by:
\begin{gather}
\eta^{\pm}(z)= \; C_{\pm}(\cos z)^{\frac{3}{2}+\tilde l} 
\hspace{12cm} \nonumber\\ \hspace{1.6cm} \times 
{}_2F_1[\frac{1}{2}(3\pm 1+\tilde l+l-\tilde\omega_{\pm}),\frac{1}{2}(3\pm 1+\tilde l+l+\tilde\omega_{\pm}),\tilde l+2,\cos^2 z](\sin z)^{\frac{3}{2}\pm 1+l}\ .
\end{gather}
Regularity of the solutions at $z=\pi$ implies the following eigenfrequencies:
\begin{equation}
\tilde\omega_{\pm}=3\pm1+l+2n_{\pm}+\tilde l;~~~~~~~n_{\pm}=0,1,2,\dots ; \ .\label{typeIIIpm}
\end{equation}  
Note that the spectrum of $\eta_+$ is contained in the spectrum of the $\eta_-$ modes. Therefore modulo the number of independent modes the spectrum of type III modes is described by:
\begin{equation}
\tilde\omega=2+l+2n+\tilde l;~~~~~~n=0,1,2,\dots ; \ .\label{typeIIIspect}
\end{equation}
On the other hand, by studying the asymptotic behaviour of the
solutions at large $\tilde\rho$, we can verify that the conformal dimension of the dual gauge invariant operators is $\Delta=3+l$. Therefore we can write our expression for the spectrum as:
\begin{equation}
\tilde\omega=\Delta-1+2n+\tilde l \ .
\end{equation}
Note that the entire spectrum is shifted by $-1$ relative
to the spectrum of type II modes. The spectrum of modes with $\tilde
l=0$ and $l=0$ requires separate analysis. Here we will simply report
the result, which is that the expression for the spectrum is still
given by equation (\ref{typeIIIspect}) at vanishing $\tilde l$ or
$l$. We will comment more on the spectrum of type III modes in the 
Section 5 below, where we reproduce our results from field theory considerations.
\subsection{Fluctuations of a D5--brane probe}
In this subsection we obtain the spectrum of fluctuations of a probe
D5--brane at vanishing bare mass $\tilde m=0$ in closed form. 
Let us first consider fluctuations of $l(r)$.
\subsubsection{Fluctuations of $l(r)$}
We consider the ansatz $\delta l=e^{i\frac{\tilde\omega}{R} t} h(\tilde r){\cal Y}^{\tilde l}(\tilde S^2){\cal Y}^l(S^2)$ for the fluctuation of the D5--brane and substitute this into the corresponding equation of motion (here $\tilde r=r/R$) . The result is a differential equation of the general form  (\ref{A1.1}) with coefficients:
\begin{gather} 
C_0(\tilde\rho)=\frac{6+8 \tilde r ^2}{\tilde r ^2-16 \tilde r
  ^6}-\frac{\tilde l(\tilde l+1)}{(\tilde
  r^2-1/4)^2}-\frac{l(l+1)}{\tilde r^2}; \nonumber\\ C_1(\tilde r)=-\frac{4+8 \tilde r^2+32 \tilde r^4}{\tilde r -16 \tilde r^5};~~~B(\tilde r)=\frac{4}{1+4\tilde r^2} \, .
\end{gather}
The calculation proceeds in complete analogy to the D3/D7 case. One applies the procedure outlined in equations (\ref{A1.1})-(\ref{Schrod}) and obtains the effective potential:
\begin{equation}
V_{\rm eff}(z)=\frac{l(l+2)}{\sin^2z}+\frac{\tilde l(\tilde
  l+2)}{\cos^2 z}
;~~~~~~~z\in\left[{\pi}/{2},\pi\right) \, ,\label{effectPotnLD5}
\end{equation}
where 
\begin{equation}
z(\tilde r)\equiv 2\arctan{(2\tilde r)}. \label{ZofRho5}
\end{equation}
The solution regular at $z=\pi/2$ is given by:
\begin{equation}
f^{\pm}(z)=C(\cos z)^{\tilde l+1}{}_2F_1[\frac{1}{2}(2+\tilde l+l-\tilde\omega),\frac{1}{2}(2+\tilde l+l+\tilde\omega),\tilde l+\frac{3}{2},\cos^2 z](\sin z)^{l+1}
\end{equation}
Regularity at $z=\pi$ requires that one of the first two arguments of
the hypergeometric function must be a non-positive integer. Without loss of generality we can consider only positive $\tilde\omega$, which implies
\begin{equation}
\frac{1}{2}(2+\tilde l+l-\tilde\omega)=-n;~~~~~~~~~n=0,1,2,\dots\ .
\end{equation}
Therefore we obtain the following expression for the spectrum:
\begin{equation}
\tilde\omega=2+l+2n+\tilde l;~~~~~~~~~n=0,1,2,\dots\ .\label{Spectm0lD5}
\end{equation}
Equation (\ref{Spectm0lD5}) is equivalent to equation (\ref{SpectSug}) with $\Delta=2,l=0,\tilde l=0$. In section 5 we will obtain this equation from field theory considerations. Now let us focus on the fluctuation of the Neumann-Dirichlet coordinate.
\subsubsection{Fluctuation of the Neumann-Dirichlet coordinate. }
In this subsection we analyze the spectrum of fluctuations along the
polar coordinate $\alpha$ defined in \eqref{parmD5} 
of the three-sphere inside the global AdS$_5$ space-time. Note that in the limit of infinite radius of the three-sphere, we obtain AdS$_5$ space-time in a Poincar\'e patch. In this limit we have an underlying description in terms of a stack of $N_c$ coincident D3--branes and the notion of Neumann-Dirichlet coordinate is well-defined. 

To obtain the spectrum of fluctuations we follow the same prescription that we did so far. We substitute $\delta\alpha=e^{\frac{\tilde\omega}{R}t}h(\tilde r)$ into the equation of motion for $\delta\alpha$ (here $\tilde r=r/R$) and obtain a differential equation of the general form (\ref{A1.1}) with coefficients:
\begin{gather} 
C_0(\tilde r)=\frac{32}{(1-4\tilde r^2)^2};~~~C_1(\tilde
r)=\frac{4+24\tilde r^2+96\tilde r^4}{\tilde r(16\tilde
  r^4-1)};~~~B(\tilde r)=\frac{4}{1+4\tilde r^2} \, .
\end{gather}
The corresponding effective potential $V_{\rm eff}(\tilde r)$ defined in equation (\ref{effpotRho}) is given by
\begin{equation}
V_{\rm eff}(\tilde r)=\frac{3(1+4\tilde r^2)^2}{8\tilde r^2}\ .
\end{equation}
After changing variables to $u(\tilde r)\equiv 2\arctan{(2\tilde r)}$ the equation of motion takes the general form (\ref{Schrod}) with effective potential $V_{\rm eff}(u)$ given by
\begin{equation}
V_{\rm eff}(u)=\frac{6}{\sin^2 u};~~~~~~u\in(\pi/2,\pi) \, .
\end{equation}
The general solution is given by:
\begin{equation}
f(u)=\left[D_1P_{\tilde\omega-1/2}^{5/2}(\cos u)+D_2Q_{\tilde\omega-1/2}^{5/2}(\cos u)\right]{\sqrt{\sin u}}\ .
\end{equation}
In the original variables ($h(\tilde\rho)$), we can verify that regularity of the solution requires that $f(\pi/2)=f(\pi)=0$. This requires:
\begin{eqnarray}
\left[\begin{matrix} 
\cos(\tilde\omega\frac{\pi}{2}) & -\frac{\pi}{2}\sin(\tilde\omega\frac{\pi}{2}) \\
\cos(\tilde\omega\pi) & -\frac{\pi}{2}\sin(\tilde\omega\pi)\\
\end{matrix}\right]
\left[\begin{matrix}D_1\\D_2\\ \end{matrix}\right]=0\ .\label{constrND}
\end{eqnarray}
Non-trivial solutions exist only if the determinant in
(\ref{constrND}) vanishes. This is equivalent to
$\sin{(\tilde\omega\frac{\pi}{2})}=0$ and hence $\tilde\omega=2n_1$,
where $n_1=0,1,2,\dots\ .$ In fact, even with this restriction we  can
verify that $h(\tilde r)$ diverges at $\tilde r=1/2$. In particular,
it behaves as $\propto1/(\tilde r-1/2)$. However this still
corresponds to bounded fluctuation of the probe D5--brane since the invariant quantity $G_{\alpha\alpha}(\delta\alpha)^2 \propto (\tilde r-1/2)^2h(\tilde r)^2<\infty$ remains bounded. Next we impose the condition that at infinity ($\tilde r\to\infty$) $h(\tilde r)$ falls as $\propto1/\tilde r^5$.~This is related to the fact that the corresponding gauge invariant operator has conformal dimension four ($\Delta=4$).~One can verify that all solutions with $n_1\geq 2$ have that asymptotic behaviour. Therefore the final expression for the spectrum is:
\begin{equation}  
\tilde\omega=4+2n;~~~~~n=0,1,2,\dots\ ,
\end{equation}
which is exactly equation (\ref{SpectSug}) with ($\Delta=4$).

\section{Spectrum at zero bare mass -- Field theory side }
\label{sec:fieldtheory}

We now turn to the analysis of the spectrum at zero bare quark mass on
the quantum field theory side of the correspondence. It is instructive
to begin with a brief review of the properties of a conformally coupled free scalar on $\IR^1\times\tilde S^n$.
\subsection{A Conformally coupled scalar on $\IR^1\times \tilde S^n$}
We begin by reviewing 
the derivation of the spectrum of a conformally coupled free scalar field on an Einstein universe $\IR^1\times \tilde S^n$. We start with the Lagrangian
\begin{equation}
{\cal L}=\sqrt{-g_{(n)}}\left[\frac{1}{2}g_{(n)}^{ab}\partial_a\phi\partial_b\phi+\frac{1}{2}\xi{\cal R}^{(n)}\phi^2\right]\ .
\end{equation}
Here ${g_{(n)}}_{ab}$ is the metric on $\IR^1\times \tilde S^n$ given by:
\begin{equation}
ds^2=-dt^2+R_n^2d\Omega_n^2\ ,\label{einn}
\end{equation}
${\cal R}^{(n)}$ is the Ricci scalar curvature of the metric in equation (\ref{einn})  given by:
\begin{equation}
{\cal R}^{(n)}=n(n-1)\frac{1}{R_n^2}
\end{equation}
and $R_n$ is the radius of $\tilde S^n$. For $\xi=(n-1)/{4(n)}$,
the scalar
is conformally coupled. Now it is straightforward to write down the equation of motion:
\begin{equation}
-\partial_0^2\phi+\frac{1}{R_n^2}\Delta_{\Omega_{n}}\phi-\frac{(n-1)^2}{4}\frac{1}{R_n^2}\phi=0\ .\label{eqmsn}
\end{equation}
Next we Fourier transform :
\begin{equation}
\phi(t,\Omega_{n})=\int d\omega \sum_{\tilde l} C_{\tilde l}(\omega)e^{i\omega t}{\cal Y}^{\tilde l}(\Omega_n);~~~\tilde l=0,1,\dots;\label{EQphiSphere}
\end{equation}
where ${\cal Y}^{\tilde l}(\Omega_n)$ are the scalar spherical harmonics on the unit $S^n$ satisfying:
\begin{equation}
\Delta_{\Omega_n}{\cal Y}^{\tilde l}(\Omega_n)=-\tilde l(\tilde l+n-1){\cal Y}^{\tilde l}(\Omega_n);~~~\tilde l=0,1,\dots;\ .
\end{equation}
The equation of motion (\ref{eqmsn}) becomes:
\begin{equation}
\left[\omega^2-\left((n-1)/2+\tilde l\right)^2\frac{1}{R_n^2}\right]C_{\tilde l}(\omega)=0\ ,
\end{equation}
which implies:
\begin{equation}
C_{\tilde l}(\omega)\propto \delta{(\omega^2-E_{\tilde l}^2)} \, , \label{Ctilde}
\end{equation}
where $E_{\tilde l}$ is given by:
\begin{equation}
E_{\tilde l}=\left((n-1)/2+\tilde l\right)\frac{1}{R_n};~~~\tilde l=0,1,\dots; \ .
\end{equation}
Now one can write equation (\ref{EQphiSphere}) as:
\begin{equation}
\phi(t,\Omega_n)=\sum_{\tilde l} \frac{1}{\sqrt{E_{\tilde l}}}\left(e^{iE_{\tilde l}t}a_{\tilde l+}\bar {\cal Y}^{\tilde l}(\Omega_n)+e^{-iE_{\tilde l}t}a_{\tilde l-}{\cal Y}^{\tilde l}(\Omega_n)\right)
\end{equation}
and proceed with quantization of $\phi$ in the standard way.

\subsection{A collection of fields on $\tilde S^3$}
In this section we calculate the spectrum of a collection of free scalar fields propagating on $S^3$. In particular we consider the field content of the gauge invariant operators holographically dual to the normal modes of the D7--brane that we computed in section~4. The flavoured dual field theory in question is a ${\cal N}=4$ SYM theory in $1+3$--dimensions coupled to an ${\cal N}=2$ hypermultiplet. The coupling of theory is conveniently studied if we decompose the ${\cal N}=4$ supermultiplet of the adjoint degrees of freedom into one vector multiplet and two ${\cal N}=2$ hypermultiplets. The bosonic content of the ${\cal N}=4$ multiplet is the $(1+3)$--dimensional vector $A_{\mu}$ and the six adjoint scalars $X^4,\dots X^9$. The bosonic sector of the vector multiplet contains the vector $A_{\mu}$ and the adjoint scalars $X_{V}^A=(X^8,X^9)$. The rest four adjoint scalars $X^i$~~~($i=4,5,6,7$) form the bosonic content of the two adjoint hypermultiplets. The fundamental fields form a hypermultiplet comprised of two spinor fields $\psi_i$ and two complex scalar fields $q^m$~~$(i,m=1,2)$. 

The operators corresponding to mesonic excitations of flavoured ${\cal
  N}=2$ SYM theory 
on flat Minkowski spacetime 
have been identified in refs. \cite{Kruczenski:2003be},\cite{Erdmenger:2007cm}. In particular the meson states have been classified in multiplets forming irreducible representations of the global $SO(4)\approx SU_{\Phi}(2)\times SU_{R}(2)$ symmetry. Furthermore it has been verified that states from the same supermultiplet (sharing the same charge under $SU_{R}(2)$ ) have the same mass spectrum. The latter has beeg determined via holographic techniques. 

The flavoured SYM theory on $\IR^1\times \tilde S^3$  that we consider
has the same global $SO(4)\approx SU_{\Phi}(2)\times SU_{R}(2)$
symmetry.  This is why we may 
use the same classification of the mesonic excitations. However, 
the supersymmetry algebra on $\IR^1\times \tilde S^3$ mixes states at
different floors of the Kaluza-Klein tower on $\tilde S^3$. Therefore
we cannot expect to have the same degeneracy of the spectrum of meson
excitations as for the theory on flat Minkowski spacetime. 
A summary of our results is provided in table \ref{tab:1}.
\begin{table}[h]
\begin{center}
\begin{tabular}{|c|c|c|c|c|c|c|}
\hline
      &\!\! fluctuation & d.o.f. & $(j_\Phi,j_{R})$  & op. & $\Delta$&spectrum\\
\hline  \vspace{-0.3cm}&&&&&&\\
mesons & 1 scalar  &1& ($\frac{\ell-1}{2}$, $\frac{\ell+1}{2}$)$$ & ${\cal C}^{I\ell}$ & $1+\ell$&$\Delta+2n+\tilde\ell$ \\
(bosons)       & 2 scalars &2& ($\frac{\ell}{2}$, $\frac{\ell}{2}$)$$  & ${\cal M}_s^{A\ell}$  & $3+\ell$&$\Delta+2n+\tilde\ell$\\
       & 1 scalar  &1& ($\frac{\ell}{2}$, $\frac{\ell}{2}$)$$ & ${\cal J}^{5\ell}$ & $3+\ell$&$\Delta+1+2n+\tilde\ell$ \\
  & 1 scalar &1 & ($\frac{\ell}{2}$, $\frac{\ell}{2}$)$$ & ${\cal J}^{0\ell}$  & $3+\ell$&$\Delta-1+2n+\tilde\ell$ \\
 & 1 tr. vector on $\tilde S^3$ &2 & ($\frac{\ell}{2}$, $\frac{\ell}{2}$)$$ & ${\cal J}^{\alpha\ell}$  & $3+\ell$&$\Delta+2n+\tilde\ell$ \\
       & 1 scalar  &1& ($\frac{\ell+1}{2}$, $\frac{\ell-1}{2}$)$$& -- & $5+\ell$ &$\Delta+2n+\tilde\ell$ \\
\hline
\end{tabular}
\end{center}
\caption{\small Bosonic sector of mesonic excitations of conformally
  coupled flavoured ${\cal N}=2$ SYM on $\IR^1\times \tilde S^3$\
  . Here  $(j_\Phi,j_{R})$ labels irreducible representations of the
  global $SO(4)\approx SU_{\Phi}(2)\times SU_{R}(2)$ symmetry, while
  $\tilde\ell$ labels the multipole momentum on $\tilde S^3$\ . Note
  also that the transverse vector ${\cal J}^{\alpha\ell}$ satisfies
  $\tilde\nabla_{\alpha} {\cal J}^{\alpha\ell}=0$ (Coulomb gauge),
  where $\tilde\nabla_{\alpha}$ is a covariant derivative on $\tilde
  S^3$. The hyphen in the fifth column corresponds to a descendent of
  the operator ${\cal C}^{I\ell}$. Note also that the spectrum of the
  pseudo-scalar ${\cal J}^{5\ell}$ and that of the descendent of
  ${\cal C}^{I\ell}$ has been obtained only from supergravity in this
  paper, while all others are obtained on both sides of the correspondence.}
\label{tab:1}
\end{table}
To begin with, let us consider the operator dual to fluctuations of
the transverse scalar modes of the probe brane (fluctuations along $L$ and $\phi$). The corresponding operator is given by \cite{Erdmenger:2007cm}:
\begin{equation}
{\cal M}_s^{Al}=\bar\psi_i\sigma_{ij}^A \chi^l\psi_j+\bar q_m X^{A}_V
\chi^l q_m~~~~(i, m = 1,2)\label{scalop} \, ,
\end{equation}
which has conformal dimension $\Delta=l+3$. Here $\chi^l$ denotes the symmetric traceless operator insertion $X^{\{i_1}\dots X^{i_l\}}$ of $l$ adjoint scalars $(i=4,5,6,7)$, $X^A_V$ denotes the vector $(X^8,X^9)$ and $\sigma^A=(\sigma^1,\sigma^2)$ is a doublet of Pauli matrices. Let us focus on the scalar part of the operator (\ref{scalop}). Without loss of generality we can consider only the $(m=1, A=8)$ additive term. In view of the result obtained in the previous subsection we expand:
\begin{eqnarray}
X^{(C)}_{ij}(t)&=&\sqrt{\frac{R_3}{2}}\left(e^{i\frac{t}{R_3}}{X^C_{0}}_{ij}+e^{-i\frac{t}{R_3}}{X^C_0}_{ij}^{\dagger}\right);~~~~C=4,5,6,7,8;\label{collection}\\
{q^i_1}^{\dagger}(t,\tilde S^3)&=&\sum_{\tilde l_1=0}^{\infty}\sum_{I_1=0}^{(\tilde l_1+1)^2}\frac{\sqrt{R_3}}{\sqrt{2+2\tilde l_1}}\left(e^{i(1+\tilde l_1)\frac{t}{R_3}}a^i_{\tilde l_1I_1}\bar {\cal Y}^{\tilde l_1I_1}(\tilde S^3)+e^{-i(1+\tilde l_1)\frac{t}{R_3}}
{b^i}^{\dagger}_{\tilde l_1I_1}{\cal Y}^{\tilde l_1I_1}(\tilde S^3)\right); \nonumber\\
q^i_1(t,\tilde S^3)&=&\sum_{\tilde l_2=0}^{\infty}\sum_{I_2=0}^{(\tilde l_2+1)^2}\frac{\sqrt{R_3}}{\sqrt{2+2\tilde l_2}}\left(e^{i(1+\tilde l_2)\frac{t}{R_3}}b^i_{\tilde l_2I_2}\bar {\cal Y}^{\tilde l_2I_2}(\tilde S^3)+e^{-i(1+\tilde l_1)\frac{t}{R_3}}{a^i}^{\dagger}_{\tilde l_2I_2}{\cal Y}^{\tilde l_2I_2}(\tilde S^3)\right)\nonumber \, .
\end{eqnarray}
Here we have written explicitly the  $i,j=1\dots N_c$ color indices. Note that we have turned on only the ``zero" mode of the adjoint scalar which is due to the conformal coupling. Next we define a vacuum state $|0\rangle$ satisfying:
\begin{equation}
a^i_{l_1I_1}|0\rangle=b^i_{l_2I_2}|0\rangle={X^C_0}_{ij}|0\rangle=0 \ .
\end{equation}
The state representing the collection of fields corresponding to the
$(m=1,A=7)$ 
component of the operator (\ref{scalop}) is given by
\begin{equation}
\bar q_1 X^{8}_V \chi^l q_1|0\rangle=\frac{R_3}{2}\sum_{\tilde l_1,\tilde l_2=0}^{\infty}\sum_{I_1=0}^{(\tilde l_1+1)^2}\sum_{I_2=0}^{(\tilde l_2+1)^2}e^{i(3+l+\tilde l_1+\tilde l_2)\frac{t}{R_3}}\frac{{\cal Y}^{\tilde l_2I_2}(\tilde S^3){\cal Y}^{\tilde l_1I_1}(\tilde S^3)}{\sqrt{(\tilde l_2+1)(\tilde l_1+1)}}a^{\dagger}_{\tilde l_2I_2}{X^8_0}^{\dagger}{\chi_0^l}^{\dagger}b^{\dagger}_{\tilde l_1I_1}|0\rangle \, .\label{expsphr}
\end{equation}
Note that the state (\ref{expsphr}) is a superposition of states with
definite energy $E_{l,J}=(3+l+J)/R_3$, where $J=\tilde l_1+\tilde
l_2$. Apparently there is a large degeneracy of the spectrum
corresponding to the different choices of $\tilde l_1$ and $\tilde
l_2$ which sum up to the same number $J$. Note also that a state with
a definite energy has a definite eigenvalue under the antipodal map
(which for spheres in even dimensions coincides with parity) on the three-sphere given by $(-1)^J$. Our next step is to expand the product of the spherical harmonics in (\ref{expsphr}) in Laplace series:
\begin{equation}
{\cal Y}^{\tilde l_1I_1}(\tilde S^3){\cal Y}^{\tilde l_2I_2}(\tilde
S^3)=\sum_{\tilde l=0}^{\infty}\sum_{I=0}^{(\tilde l+1)^2}{\bf
  C}^{\tilde lI}_{\tilde l_1I_1,\tilde l_2I_2}{\cal Y}^{\tilde l
  I}(\tilde S^3)\,  . \label{CL-GR}
\end{equation}
The coefficients ${\bf C}^{\tilde lI}_{\tilde l_1I_1,\tilde l_2I_2}$
are non-zero only for $|\tilde l_1-\tilde l_2 | \leq \tilde l\leq
\tilde l_1+\tilde l_2$ (addition of angular momentum) and
$(-1)^{\tilde l}=(-1)^{\tilde l_1+\tilde l_2}$(conservation of the
antipodal map eigenvalue)\footnote {We refer the reader to the
  appendix of ref.~\cite{Hamada:2003jc} for a detailed review of the
  properties of $SU(2)\times SU(2)$ Clebsch-Gordan coefficients. Note
  that half integer quantum numbers rather than integer ones are used
  in ref.~\cite{Hamada:2003jc}.}. Therefore the Laplace series in
(\ref{CL-GR}) terminates at $J=\tilde l_1+\tilde l_2$.
For all values of $\tilde l$ that appear in the expansion, we can write $J=2n+\tilde l$ for some non-negative integer $n$. This implies that a state with a definite energy $E_{J,l}$ can be expanded as:
\begin{equation}
|E_{J,l}\rangle=\frac{R_3}{2}e^{iE_{J,l}t}\sum_{2n+\tilde l=J}\sum_{I=0}^{(\tilde l+1)^2}{\cal Y}^{\tilde l I}(\tilde S^3)C_{n\tilde l I}^{\dagger}|0\rangle\ ,
\end{equation}
where $C_{n\tilde l I}^{\dagger}$ is defined by:
\begin{equation}
C_{n\tilde l I}^{\dagger}\equiv\sum_{\tilde l_1+\tilde l_2=2n+\tilde
  l}\sum_{I_1=0}^{(\tilde l_1+1)^2}\sum_{I_2=0}^{(\tilde
  l_2+1)^2}\frac{{\bf C}^{\tilde lI}_{\tilde l_1I_1,\tilde
    l_2I_2}}{\sqrt{(\tilde l_2+1)(\tilde l_1+1)}}a^{\dagger}_{\tilde
  l_2I_2}{X^8_0}^{\dagger}{\chi_0^l}^{\dagger}b^{\dagger}_{\tilde
  l_1I_1} \, .
\end{equation}
Now the general state (\ref{expsphr}) can be written as:
\begin{equation}
\bar q_1 X^{8}_V \chi^l q_1|0\rangle=\frac{R_3}{2}\sum_{n,\tilde l=0}^{\infty}\sum_{I=0}^{(\tilde l+1)^2}e^{i(3+l+2n+\tilde l)\frac{t}{R_3}}{\cal Y}^{\tilde l I}(\tilde S^3)C_{n\tilde l I}^{\dagger}|0\rangle\ .\label{collectspect}
\end{equation}
Therefore if we define $\tilde\omega=\omega R_3$, 
our final expression for the dispersion relation is given by 
\begin{equation}
\tilde\omega=3+l+2n+\tilde l;~~~~~~~~~n=0,1,2,\dots\ .\label{SpectrumD7m0FT}
\end{equation}
This coincides
precisely with equation (\ref{SpectrumD7m0}) which is obtained for the
dual mode on the gravity side of the correspondence. 
The fact that our free field theory analysis reproduces the spectrum
at strong coupling 
is consistent with the non-renormalization theorem for chiral
primaries. 
Note that this matching provides a non-trivial check of the holographic  technique employed in Section~4.

{\bf Vector modes.} Next we apply the same approach to study the spectrum of gauge invariant operators corresponding to fluctuations of the $U(1)$ worldvolume gauge field of the D7--brane. There is only one chiral primary operator corresponding to the type I $A_i^{-}$ mode analyzed in section 4. It  has conformal dimension $\Delta_-=1+l$ and is given by\footnote{Note that we are using the convention for the quantum number $l$ employed in ref.~\cite{Kruczenski:2003be}, namely the ground state has $l=1$.}:
\begin{equation}
{\cal C}^{Il}=\bar q^m\sigma_{mn}^I\chi^{l-1}q^n\ ,
\end{equation}
where $\sigma^I_{mn}$ $(I=1,2,3)$ is a triplet of Pauli matrices. To obtain the spectrum of the corresponding collection of scalar fields one may apply a procedure analogous to the one outlined in equations (\ref{collection})-(\ref{collectspect}). The ground state is again given by the sum of the zero point energies of the constituent fields, namely $(1+1+(l-1))/R_3=(1+l)/R_3$. Furthermore by expanding in Laplace series and employing properties of the Clebsch-Gordan coefficients on can argue that the spectrum is given by the negative sign in equation (\ref{SPECTRUMPM}):
\begin{equation}
\tilde\omega^{-}=\Delta_{-}+2n+\tilde l~~~~~~~~~n=0,1,2,\dots\ .\label{SPECTRUMM}
\end{equation}
The operator corresponding to the $A_i^+$ mode is descendent of the operator ${\cal C}^{Il}$ and can be analyzed in a similar way. 

Next we consider the spectrum of type II modes described in section 4. Note that our type II modes are subset of the type II modes defined in refs.~\cite{Kruczenski:2003be}, \cite{Erdmenger:2007cm}. The latter correspond to the components of the $U(N_f)$ flavour current operator \cite{Erdmenger:2007cm} in $1+3$ dimensions:
\begin{equation}
{\cal J}^{\mu l}=\bar\psi_i^a\gamma^{\mu}_{ab}\chi^l\psi_i^b+i\bar q_m\chi^lD^{\mu}q_m-i\bar D^{\mu}\bar q_m\chi^lq_m~~~~(\mu=0,1,2,3)\label{FLcurrent}
\end{equation}
and satisfy $\partial_{\mu}{\cal J}^{\mu l}=0$ (Lorentz gauge) which
leaves only three independent degrees of freedom. In our case the
field theory is defined on 
$\mathbbm{R}^1\times\tilde S^3$ and the Lorentz symmetry is broken. Our type II modes (look at section 4) comprise of the spacial part of the current operator (\ref{FLcurrent}) defined on $\tilde S^3$ and satisfy $\tilde\nabla_{\alpha}{\cal J}^{\alpha I}=0$ (Coulomb gauge). Therefore we have only two independent degrees of freedom which have Laplace expansion in terms of  transverse vector spherical harmonics ${\cal Y}^{\tilde l\pm}_{\alpha}$. It turns out that one can successfully apply the procedure outlined in (\ref{collection})-(\ref{collectspect}) with minor modifications. Consider the scalar part of the operator (\ref{FLcurrent}) one can see that the energy levels of the corresponding collection of fields are given by $(1+1+l+\tilde l_1+\tilde l_1)/R_3$. The relevant Laplace expansion is given by:
\begin{equation}
{\cal Y}^{\tilde l_1I_1}(\tilde S^3)\tilde\nabla_{\alpha}{\cal Y}^{\tilde l_2I_2}(\tilde S^3)=\sum_{\tilde l=|\tilde l_1-\tilde l_2|+1}^{\tilde l_1+\tilde l_2-1}\sum_{I=0}^{(\tilde l+1)^2}{\bf G}^{lI}_{\tilde l_1I_1,\tilde l_2I_2}{\cal Y}_{\alpha}^{\tilde l\pm, I}(\tilde S^3)\, . \label{CL-GG}
\end{equation}
Note that the coefficients ${\bf G}^{\tilde lI}_{\tilde l_1I_1\tilde
  l_2I_2}$ are non vanishing only for  $\tilde l_1+\tilde
l_2=2n+1+\tilde l$ \cite{Hamada:2003jc} for non-negative integer
values of $n$. Therefore the final expression for the spectrum of
$\tilde\omega$ is given by equation (\ref{SpectrumD7m0FT}) again.

Finally we consider the spectrum of type III modes. The corresponding operators are the time component of the $U(N_f)$ flavour current (\ref{FLcurrent}) and the pseudo-scalar operator \cite{Erdmenger:2007cm}:
\begin{equation}
{\cal J}^{5l-1}=\bar\psi^a_i\gamma^5_{ab}\chi^{l-1}\psi_i^b+\dots ;~~~~(l\geq 1)\ .\label{ps-scalar}
\end{equation}
There are is also a contribution from the space-like components of the
flavour current (\ref{FLcurrent}) which have expansion in longtudinal
vector spherical harmonics $\tilde\nabla_\alpha{\cal Y}^{\tilde
  l}(\tilde S^3)$. The corresponding spectrum (modulo degeneracy of
the ground state) was calculated in section 4 and is given by equation
(\ref{typeIIIspect}). To obtain the spectrum from field theory
considerations, we consider the relevant contributions from the scalar
fields to
the flavour current (\ref{FLcurrent}). The energy levels of the
corresponding collection of fields is given by $(1+1+l+\tilde
l_1+\tilde l_1)/R_3$. The relevant Laplace expansion is in terms of
scalar spherical harmonics (for the time-like component of the flavour
current) and longitudinal vector spherical harmonics
$\tilde\nabla_\alpha{\cal Y}^{\tilde l}$ (for the space-like
components of the flavour current). Using properties of the $C^{\tilde
  lI}_{\tilde l_1I_1\tilde l_2I_2}$ Clebsch-Gordan coefficients, we
can show that in both cases the $\tilde l$-th terms in the expansion
vanish unless $\tilde l_1+\tilde l_2=2n+\tilde l$. Therefore the
spectrum is given by 
\begin{equation}
\tilde\omega=2+l+2n+\tilde l\ , \label{fieldIII}
\end{equation}
which coincides with the gravity result  (\ref{typeIIIspect}). Note that the type III vector modes analyzed in section \ref{sec:gravity} have two independent modes $\eta_{\pm}$ with spectrum $\tilde\omega_{\pm}$ given by equation (\ref{typeIIIpm}):
\begin{equation}
\tilde\omega_{\pm}=3\pm1+l+2n+\tilde l\ , \label{fieldIIIpm}
\end{equation}
and we verified that the contributions from the scalar fields to 
the flavour current (\ref{FLcurrent}) reproduce the spectrum of the $\eta_{-}$ modes. It is natural to assume that the pseudo-scalar operator (\ref{ps-scalar}) corresponds to the $\eta_+$ modes and has spectrum given by the positive sign in equation (\ref{fieldIIIpm}).

This completes our study of the meson spectrum of the bosonic sector of conformally coupled flavoured ${\cal N}=2$ SYM theory on $\IR^1\times \tilde S^3$.
\subsection{A collection of fields on $\tilde S^2$}\vspace{-.29cm} \vspace{.29cm} 
In this subsection we consider the gauge field theory holographically dual to a D5--brane probe in a global AdS$_5\times S^5$ space-time. 
The theory in question is an ${\cal N}=4$ SYM theory on $\IR^1\times
\tilde S^3$ coupled to a defect field theory defined on an equatorial
$\tilde S^2\subset \tilde S^3$. At vanishing bare quark mass,  the
theory is superconformal  and preserves ${\cal N}=4$ supersymmetry in
2+1 dimensions. The coupling of theory is the same as the one for the
theory on flat Minkowski space studied for a first time in
refs.~\cite{{DeWolfe:2001pq}, {Erdmenger:2002ex}}  and reviewed in
ref.~\cite{Ammon:2010pg}. The analysis requires decomposing the
$(3+1)$--dimensional ${\cal N}=4$ supermultiplet into two
$(2+1)$--dimensional ${\cal N}=4$ supermultiplets, a vector multiplet
and a hypermultiplet. The bosonic content of the $(3+1)$--dimensional
${\cal N}=4$ multiplet is the $(3+1)$--dimensional vector $A_{\mu}$
and the six adjoint scalars $X^4,\dots X^9$. The bosonic content of
the $(2+1)$--dimensional vector multiplet comprises of the $2+1$
dimensional vector $A_k$~~$k=0,1,2$ and the three scalars
$X_V^A=(X^7,X^8,X^9)$. The bosonic content of the $(2+1)$--dimensional
hypermultiplet is the scalar $A_3$ and the scalars
$X_H=(X^4,X^5,X^6)$. The flavour fields form a $(2+1)$--dimensional
hypermultiplet with two fermions $\psi_i$ and two complex scalars
$q^m$ ($i,m=1,2$). \\ \indent
A full analysis of the spectrum of mesonic excitations is beyond the
scope of this paper. In this subsection we calculate the spectrum of
fluctuations of the operator corresponding to the fundamental
condensate of the defect field theory given by the operator
\cite{Ammon:2010pg}:
\begin{equation}
{\cal E}^{Al}=\bar\psi_i\sigma^{A}_{ij}\chi^l\psi_j+2\bar q^mX_V^A\chi^lq^m;~~~~ (A=1,2,3; ~~i,m=1,2)\ .\label{scalopD5}
\end{equation}
Here $\chi^l$ denotes the symmetric traceless operator insertion $X^{\{i_1}\dots X^{i_l\}}$ of $l$ adjoint scalars $(i=4,5,6)$, $X^A_V$ denotes the vector $(X^7,X^8,X^9)$ and $\sigma^A=(\sigma^1,\sigma^2,\sigma^3)$ is a triplet of Pauli matrices. On gravity side this operator is dual to the fluctuations modes of the transverse scalars of the probe D5--brane. For a particular component (say $A=7$), the operator ${\cal E}^{Al}$ corresponds to fluctuations of the profile function $l(r)$ studied in section~4. Therefore the corresponding spectrum is given by equation (\ref{Spectm0lD5}). Our goal is to obtain the same result from field theory considerations. Note that the operator ${\cal E}^{Al}$ is the analog of the operator ${\cal M}_s^{Al}$ considered in the previous subsection. This implies that the spectrum can be obtained by adapting the procedure outlined in equations  (\ref{collection})-(\ref{collectspect}) to the defect field theory defined on the equatorial $\tilde S^2$. To begin with let us focus on the scalar sector of the $A=7$ component of the operator ${\cal E}^{Al}$ and consider only the $m=1$ additive term. The analog of equation (\ref{collection}) is given by:
\begin{eqnarray}
X^{(C)}_{ij}(t)&=&\sqrt{\frac{R_3}{2}}\left(e^{i\frac{t}{R_3}}{X^C_{0}}_{ij}+e^{-i\frac{t}{R_3}}{X^C_0}_{ij}^{\dagger}\right);~~~~C=4,5,6,7;\label{collectionD5}\\
{q^i_1}^{\dagger}(t,\tilde S^3)&=&\sum_{\tilde l_1=0}^{\infty}\sum_{m_1=-\tilde l_1}^{\tilde l_1}\frac{\sqrt{R_3}}{\sqrt{1+2\tilde l_1}}\left(e^{i(\frac{1}{2}+\tilde l_1)\frac{t}{R_3}}a^i_{\tilde l_1m_1}\bar {\cal Y}^{\tilde l_1m_1}(\tilde S^2)+e^{-i(\frac{1}{2}+\tilde l_1)\frac{t}{R_3}}
{b^i}^{\dagger}_{\tilde l_1m_1}{\cal Y}^{\tilde l_1m_1}(\tilde S^2)\right); \nonumber\\
q^i_1(t,\tilde S^3)&=&\sum_{\tilde l_2=0}^{\infty}\sum_{m_2=-\tilde
  l_2}^{\tilde l_2}\frac{\sqrt{R_3}}{\sqrt{1+2\tilde
    l_2}}\left(e^{i(\frac{1}{2}+\tilde l_2)\frac{t}{R_3}}b^i_{\tilde
    l_2m_2}\bar {\cal Y}^{\tilde l_2m_2}(\tilde
  S^2)+e^{-i(\frac{1}{2}+\tilde
    l_1)\frac{t}{R_3}}{a^i}^{\dagger}_{\tilde l_2m_2}{\cal Y}^{\tilde
    l_2m_2}(\tilde S^2)\right) \nonumber \ .
\end{eqnarray}
Note that in equation (\ref{collectionD5}) the adjoint fields $X^{C}$ are truncated to the zero modes on $\tilde S^3$, while the fundamental scalars are expanded in spherical harmonics on $\tilde S^2$ and have zero point energies $1/(2R_3)$. Note also that $R_2=R_3$ since the fundamental fields propagate on an equatorial two-sphere $\tilde S^2$ inside the three-sphere $\tilde S^3$. Next we define a vacuum state $|0\rangle$ satisfying:
\begin{equation}
a^i_{l_1m_1}|0\rangle=b^i_{l_2m_2}|0\rangle={X^C_0}_{ij}|0\rangle=0 \ .
\end{equation}
The state representing the collection of fields corresponding to the $(m=1,A=7)$ component of the operator (\ref{scalopD5}) is given by:
\begin{gather}
\bar q_1 X^{7}_V \chi^l q_1|0\rangle=  \hspace{11cm} \nonumber\\ 
\hspace{2cm} \sum_{\tilde l_1,\tilde l_2=0}^{\infty}\sum_{m_1=-\tilde
  l_1}^{\tilde l_1}\sum_{m_2=-\tilde l_2}^{\tilde
  l_2}R_3e^{i(2+l+\tilde l_1+\tilde l_2)\frac{t}{R_3}}\frac{{\cal
    Y}^{\tilde l_2m_2}(\tilde S^2){\cal Y}^{\tilde l_1m_1}(\tilde
  S^2)}{\sqrt{(2\tilde l_2+1)(2\tilde l_1+1)}}a^{\dagger}_{\tilde
  l_2m_2}{X^7_0}^{\dagger}{\chi_0^l}^{\dagger}b^{\dagger}_{\tilde
  l_1m_1}|0\rangle \ . \nonumber\\ \label{expsphrD5}
\end{gather}
Note that the state (\ref{expsphrD5}) is a superposition of states with definite energy $E_{l,J}=(2+l+J)/R_3$, where $J=\tilde l_1+\tilde l_2$. Note also that a state with a definite energy has a definite parity given by $(-1)^J$. Next we expand:
\begin{equation}
{\cal Y}^{\tilde l_1m_1}(\tilde S^2){\cal Y}^{\tilde l_2m_2}(\tilde
S^2)=\sum_{\tilde l=|\tilde l_1-\tilde l_2|}^{\tilde l_1+\tilde
  l_2}\sum_{m=-\tilde l}^{\tilde l}{\bf C}^{\tilde l m}_{\tilde
  l_1m_1,\tilde l_2m_2}{\cal Y}^{\tilde l m}(\tilde S^2)\,  . \label{CL-GR-SU2}
\end{equation}
The coefficients ${\bf C}^{\tilde lm}_{\tilde l_1m_1,\tilde l_2m_2}$
are non-zero only for $|\tilde l_1-\tilde l_2 | \leq \tilde l\leq \tilde l_1+\tilde l_2$ (addition of angular momentum) and $(-1)^{\tilde l}=(-1)^{\tilde l_1+\tilde l_2}$(conservation of parity). Therefore for all values of $\tilde l$ that appear in the expansion, we can write $J=2n+\tilde l$ for some non-negative integer $n$. This implies that a state with a definite energy $E_{J,l}$ can be expanded as:
\begin{equation}
|E_{J,l}\rangle=R_3e^{iE_{J,l}t}\sum_{2n+\tilde l=J}\sum_{m=-\tilde l}^{\tilde l}{\cal Y}^{\tilde l m}(\tilde S^2)C_{n\tilde l m}^{\dagger}|0\rangle\ ,
\end{equation}
where $C_{n\tilde l m}^{\dagger}$ is defined by:
\begin{equation}
C_{n\tilde l m}^{\dagger}\equiv\sum_{\tilde l_1+\tilde l_2=2n+\tilde
  l}\sum_{m_1=-\tilde l_1}^{\tilde l_1}\sum_{m_2=-\tilde l_2}^{\tilde
  l_2}\frac{{\bf C}^{\tilde l m}_{\tilde l_1m_1,\tilde
    l_2m_2}}{\sqrt{(2\tilde l_2+1)(2\tilde l_1+1)}}a^{\dagger}_{\tilde
  l_2m_2}{X^7_0}^{\dagger}{\chi_0^l}^{\dagger}b^{\dagger}_{\tilde
  l_1m_1} \, .
\end{equation}
Now the general state (\ref{expsphrD5}) can be written as:
\begin{equation}
\bar q_1 X^{8}_V \chi^l q_1|0\rangle=\sum_{n,\tilde l=0}^{\infty}\sum_{m=-\tilde l}^{\tilde l}R_3e^{i(2+l+2n+\tilde l)\frac{t}{R_3}}{\cal Y}^{\tilde l I}(\tilde S^2)C_{n\tilde l m}^{\dagger}|0\rangle\ .\label{collectspectD5}
\end{equation}
Therefore if we define $\tilde\omega=\omega R_3$, our final expression for the dispersion relation is given precisely by equation (\ref{Spectm0lD5}):
\begin{equation}
\tilde\omega=2+l+2n+\tilde l;~~~~~~~~~n=0,1,2,\dots\ .\label{SpectrumD5m0FT}
\end{equation}
This completes our study of the meson spectrum of the defect field
theory on $\tilde S^2$. The free field theory analysis of the spectrum
agrees with the results obtained at strong coupling via holographic
techniques. Similarly to the case of the flavoured gauge field theory
on $\tilde S^3$ considered in the previous subsection, this is
in agreement with a non-renormalization theorem. 
However the perfect matching of equations  (\ref{Spectm0lD5}) and (\ref{SpectrumD5m0FT}) provides a satisfying verification of the validity of the holographic approach towards strongly coupled systems at least in the presence of some supersymmetry.

\section{Conclusion}

We have considered meson spectra on $\mathbbm{R} \times S^3$ in a
holographic approach of embedding D7 or D5 brane probes into global
AdS space. First, in a numerical approach, we determined the bare
quark mass dependence of the spectra through the phase transition
which occurs between branes which do or do not reach the $S^3$ sphere
in the field theory directions. Second, we studied the spectra at zero
bare quark mass analytically and found exact agreement between the
gravity and field theory calculations. For the field theory
calculation, we expanded the fields in the operators considered on
$S^3$ and combined them using the relevant Clebsch-Gordan coefficients
together with a generalized parity argument. The fact that the free
field calculation agrees with the gravity calculation is consistent
with the non-renormalization theorems of supersymmetric theories. 
Moreover, our calculation confirms the physical interpretation that
since the zero-point energy on $S^3$ exceeds the binding energy of the
mesons, the mesons cannot pair-produce any more. 

As far as we know, this is the first example of a calculation within
gauge/gravity duality with added flavour using the Born-Infeld action 
where there is an exact match
between a non-trivial field theory calculation and the gravity
result. In the plane wave limit, field theories similar to ours have
been studied previously in \cite{DeWolfe:2004zt,Chen:2004mu,Chen:2004yf}.  
We hope that 
the results of this paper
will inspire further field-theory calculations in top-down
gauge/gravity models where the dual field theory is known explicitly,
in view of further comparisons between weak and strong coupling
results for observables of physical relevance.

\section{Acknowledgements}

We are grateful to Tameem Albash, Andy O'Bannon, Thanh Hai Ngo and Andrey Zayakin for
discussions. We would like to thank the organizers of the ESI program
on ``AdS Holography and the Quark Gluon Plasma'' at the
Erwin-Schr\"odinger-Institute in Vienna for hospitality during the
early stages of this work. V.F.~would like to thank APCTP in Pohang
for hospitality during the focus program on ``Aspects of Holography and Gauge/string duality''. The work of V.F.~is
supported by an IRCSET INSPIRE Postdoctoral Fellowship. This work is
supported in part by the Cluster of Excellence `Origin and Structure
of the Universe'.

\end{document}